\documentclass{article}
\usepackage{amsfonts}

\newtheorem{theorem}{Theorem}
\newtheorem{acknowledgement}[theorem]{Acknowledgement}

\input{tcilatex}
\begin{document}

\title{Dynamics of Functional Phase Space Distribution in QFT: A Third
Quantization and Dynamical Unification of QFT and CMP}
\author{Felix A. Buot \\
%EndAName
C\&LB Research Institute, Carmen, Cebu 6005, Philippines, \\
LCFMNN, TCSE Group, Department of Physics, \\
University of San Carlos, Talamban, Cebu City 6000, Philippines}
\maketitle

\begin{abstract}
We proposed a third quantization scheme to derive the quantum dynamics of
the functional phase space distribution in quantum field theory (QFT). The
derivation is straightforward and algorithmic. This readily yields the
ballistic quantum transport equation of$\ $QFT distribution in $\left( 
\mathfrak{p},\mathfrak{q}\right) $- functional phase space, not in ordinary
position-momentum $\left( p,q\right) $-space. Our starting point is the
general mixed space representation in QFT. The end result serves as a
unification of the quantum superfield transport theory of condensed matter
physics (CMP) and QFT. This is summarized in a Table of correspondence. This
third quantization scheme may have significance in quantum fluctuation
theory of systems with many degrees of freedom.\ It may have relevance to
cosmology: gravity, multi-universes, and Yang-Mills theory.
\end{abstract}

\section{Introduction}

The unification of mixed representation in quantum field theory (QFT) and
condensed matter physics (CMP) has been enunciated by the authors of Ref. 
\cite{mixed}. For our present purpose, we recapitulate some of the crucial
steps in the construction of the general mixed representaion in QFT. This
will constitute our starting point in constructing the third quantization
scheme for investigating the quantum dynamics of the $Q$-distribution in QFT
that was derived in our previous paper \cite{mixed}. A serious discussion of
the $Q$-distribution in QFT is fairly recent and has been proposed in the
literature \cite{drum1,drum2,fried}, however a straightforward derivation of
ensuing quantum dynamics, as well as the construction of generating
functional for the QFT distribution has not been fully addressed, e.g, to
derive nonequilibrium quantum transport equations of the $Q$-distribution in
QFT. As it turns out a third quantization scheme is called for in
investigating the quantum dynamics of the $Q$-distribution in $\left( 
\mathfrak{p},\mathfrak{q}\right) $- functional phase space of QFT. The
results of this paper serve to unify the quantum superfield kinetic
transport theory of CMP and QFT. This is summarized in a Table of
correspondence between CMP in the usual second quantization formalism and
QFT in the third quantization scheme.

The idea of the third quantization, as promulgated by Raptis \cite{raptis},
involves theoretical independence with spacetime manifolds (which carries
all sort of ambigous singularities) in favor of an abstract superspace of
canonical functional fields. This has been heuristically discussed by Raptis 
\cite{raptis} in his proposed third quantization of gravity and Yang-Mills
theories. This means, in our case, avoiding the use of conventional position
and momentum dynamical variables, $\left( p,q\right) $-phase space, in favor
of canonical functional variables, $\left( \mathfrak{p},\mathfrak{q}\right) $%
- functional phase space, which is crucial in exploring the quantum dynamics
of the $Q$-distribution in $\left( \mathfrak{p},\mathfrak{q}\right) $-
functional phase space of QFT. This purpose of this paper is to elucidate
this point.

The spirit of the idea of third quantization proposed by Giddings and
Strominger \cite{goodstrom} somewhat affirm the thinking of Ref. \cite%
{raptis}. Our algorithmic procedure presented here follows the ideas brought
forth in Refs. \cite{raptis} and to some degree with \cite{goodstrom}.

A different approach proposed by Prosen et al. \cite{prosen,prosen2} is
based on the idea of using a Fock space of density operators of physical
states, obeying the quantum Liouville equation, and adjoint structure of
canonical creation and annihilation maps over this space, majorana operators
for fermions. They also called this third quantization, which is geared for
solving Langevin transport equation. However their theory is not based on
canonical functional variables i.e., abstract superspace of canonical
functional fields, such as $\left( \mathfrak{p},\mathfrak{q}\right) $-
functional phase space, consistent with spacetime manifolds independence
(i.e., in our case, $\left( p,q\right) $ independent). In other words, not
based on creation and annihilation of second quantized fields, but
specifically based on creation and annihilation operator maps over Fock
space of density operators of the Liouville equation, supposedly taking the
place of the Schr\"{o}dinger equation of the second quantization. We believe
the theory of Prosen et al. \cite{prosen,prosen2} should be aptly called
super-quantization, since it just mimic the conventional second quantization
scheme over the wavefunctions space of the Schr\"{o}dinger equations. Their
approach is on Fock space of density operators, which have a lattice Weyl
transform in conventional $\left( p,q\right) $-phase space. The lattice Weyl
transform of a density operator itself is the conventional Wigner
distribution function, $f_{w}\left( p,q\right) $. Thus, this is not the sort
of third quantization scheme discussed in this paper. The third quantization
in Ref. \cite{tunnel} also mimics the second quantization of QFT, as the
wave function of the spacetime and matter fields of the Wheeler--DeWitt
(WDW) equation is elevated to an operator, which can then be Fourier
expanded through the momentum of the scalar field The third quantization
scheme presented here is more in line with the theory proposed in Ref. \cite%
{raptis} and partly with the theory in Ref. \cite{goodstrom}.

\section{Recapitulation of Mixed Space Representation in QFT}

Let the operators $\hat{\psi}$ and $\hat{\psi}^{\dagger }$ be non-Hermitian
operators in the second quantization \footnote{%
In what follows, we do not adopt a different symbol between ordinary
derivatives and functional derivatives (this is understood from the context
of the formalism) in order to draw the parallelism between CMP and QFT.}. \
These operators obey either communtation relation for bosons or
anti-commutation relations for fermions, i.e.,%
\begin{equation}
\left[ \hat{\psi},\hat{\psi}^{\dagger }\right] _{\eta }=1  \label{one39}
\end{equation}%
where the subscript $\eta =$$+$ is for the anti-commutation and $\eta =$$-$
stands for commutation relation. These non-Hermitian operators have
distinguished left and right eigenvectors. We have%
\begin{eqnarray}
\hat{\psi}\left\vert \alpha \right\rangle &=&\alpha \left\vert \alpha
\right\rangle  \label{one40} \\
\left\langle \alpha \right\vert \hat{\psi}^{\dagger } &=&\left\langle \alpha
\right\vert \alpha ^{\ast }  \label{one41}
\end{eqnarray}%
\begin{eqnarray}
\hat{\psi}^{\dagger }\left\vert \beta \right\rangle &=&\beta \left\vert
\beta \right\rangle  \label{one42} \\
\left\langle \beta \right\vert \hat{\psi} &=&\left\langle \beta \right\vert
\beta ^{\ast }  \label{one43}
\end{eqnarray}%
This means the left eigenvector of $\hat{\psi}$ is $\left\langle \beta
\right\vert $, with eigenvalue $\beta ^{\ast }$, whereas the left
eigenvector of $\hat{\psi}^{\dagger }$ is $\left\langle \alpha \right\vert $%
, with eigenvalue $\alpha ^{\ast }$. We have,%
\begin{eqnarray}
\left\langle \beta \right\vert \hat{\psi}\left\vert \alpha \right\rangle
&=&\alpha \left\langle \beta \right\vert \left\vert \alpha \right\rangle 
\nonumber \\
&=&\beta ^{\ast }\left\langle \beta \right\vert \left\vert \alpha
\right\rangle  \nonumber \\
\left( \alpha -\beta ^{\ast }\right) \left\langle \beta \right\vert
\left\vert \alpha \right\rangle &=&0  \label{one44}
\end{eqnarray}%
and similarly%
\begin{eqnarray}
\left\langle \alpha \right\vert \hat{\psi}^{\dagger }\left\vert \beta
\right\rangle &=&\alpha ^{\ast }\left\langle \alpha \right\vert \left\vert
\beta \right\rangle  \nonumber \\
&=&\beta \left\langle \alpha \right\vert \left\vert \beta \right\rangle 
\nonumber \\
\left( \alpha ^{\ast }-\beta \right) \left\langle \alpha \right\vert
\left\vert \beta \right\rangle &=&0  \label{one45}
\end{eqnarray}%
From Eqs. (\ref{one44}) and (\ref{one45}), if $\alpha _{n}\neq \beta
_{m}^{\ast }$, $\left\langle \beta _{m}\right\vert \left\vert \alpha
_{n}\right\rangle =0$, imitating orthogonal Hermitian Hilbert space.
However, if $\alpha _{n}=\beta _{m}^{\ast }$, then $\left\langle \beta
_{m}\right\vert \left\vert \alpha _{n}\right\rangle \neq 0$, amenable to
probabilistic interpretation. Moreover, with $\left\vert \alpha
\right\rangle $ and $\left\langle \beta \right\vert $ we have the general
projection given by%
\begin{equation}
\sum\limits_{n}\frac{\left\vert \alpha _{n}\right\rangle \left\langle \beta
_{n}\right\vert }{\left\langle \beta _{n}\right\vert \left\vert \alpha
_{n}\right\rangle }=1  \label{one46}
\end{equation}%
which project only within the paired eigenspace of $\left\{ \left\vert
\alpha \right\rangle ,\left\langle \beta \right\vert \right\} $. Similarly, 
\[
\sum\limits_{n}\frac{\left\vert \beta _{n}\right\rangle \left\langle \alpha
_{n}\right\vert }{\left\langle \alpha _{n}\right\vert \left\vert \beta
_{n}\right\rangle }=1 
\]%
only projects within the paired space, $\left\{ \left\langle \alpha
\right\vert ,\left\vert \beta \right\rangle \right\} $.

The general proof of Eq. (\ref{one46}) lies in the following expansion of $%
\left\vert \Psi \right\rangle $, where $\left\vert \Psi \right\rangle $ is a
complete orthonormal eigenvector. Let, 
\[
\left\vert \Psi \right\rangle =\sum\limits_{n^{\prime }}c^{n^{\prime
}}\left\vert \alpha _{n^{\prime }}\right\rangle . 
\]%
Then we have%
\begin{eqnarray*}
\sum\limits_{n}\frac{\left\vert \alpha _{n}\right\rangle \left\langle \beta
_{n}\right\vert \left\vert \Psi \right\rangle }{\left\langle \beta
_{n}\right\vert \left\vert \alpha _{n}\right\rangle } &=&\sum\limits_{n,n^{%
\prime }}\frac{\left\vert \alpha _{n}\right\rangle \left\langle \beta
_{n}\right\vert \left\vert \alpha _{n^{\prime }}\right\rangle c^{n^{\prime }}%
}{\left\langle \beta _{n}\right\vert \left\vert \alpha _{n}\right\rangle } \\
&=&\sum\limits_{n}c^{n}\left\vert \alpha _{n}\right\rangle =\left\vert \Psi
\right\rangle
\end{eqnarray*}

\section{\label{dual_H}Mixed Hilbert-Space Construction}

Thus, from Eqs. (\ref{one44}) and (\ref{one45}), we have a well-defined
Hermitian-like operation in terms of the paired eigenvector set $\left\{
\left\langle \alpha \right\vert ,\left\vert \beta \right\rangle \right\} $
and$\left\{ \left\langle \beta \right\vert ,\left\vert \alpha \right\rangle
\right\} $ as dual eigenspaces. Indeed, assuming nondegenerates countable
states, we can pair the states so that $\alpha _{n}=\beta _{m}^{\ast }$ and
rewrite the pairs with same quantum subscript, i.e., $\alpha _{n}=\beta
_{n}^{\ast }$, which would be consistent with the commutation relation, $%
\left[ \hat{\psi}_{n},\hat{\psi}_{n}^{\dagger }\right] _{\eta }=1$.

Just as the dot product $\left\langle p\right\vert \left\vert q\right\rangle
\neq 0$ in condensed matter physics discussions, so $\left\langle \alpha
_{n}\right\vert \left\vert \alpha _{n}\right\rangle \neq 0$, which means
that $\left\langle \alpha _{n}\right\vert \left\vert \alpha
_{n}\right\rangle $ cannot be made equal to zero, since these belongs to
separate paired set, $\left\{ \left\langle \alpha \right\vert ,\left\vert
\beta \right\rangle \right\} $ and$\left\{ \left\langle \beta \right\vert
,\left\vert \alpha \right\rangle \right\} $, respectively. In some sense,
the $\left\vert \alpha \right\rangle $,$\left\vert \beta \right\rangle $
respective spaces are reminiscent of the $q$-$p$ phase space, where $%
\left\langle p\right\vert \left\vert q\right\rangle \neq 0$ is the
transition function. This conclusion can be made quite general in what
follows.

\section{\label{pairing}Pairing Algorithm and Hermitianization}

Thus, in order to work with Hermitian-like operators, we want the
eigenvalues $\alpha =\beta ^{\ast }$ and $\alpha ^{\ast }=\beta $. For
nondegenerate countable finite system, this type of pairing of the different 
$\left\vert \alpha \right\rangle $ and $\left\vert \beta \right\rangle $
spaces is well defined. For convenience in what follows, we relabel the $%
\left\vert \alpha \right\rangle $ and $\left\vert \beta \right\rangle $
notations and their adjoints to reflect the Hermitian-like new spaces, sort
of \textit{renormalize} new dynamical vector spaces.

From Eqs. (\ref{one44}) and (\ref{one45}), it seems trivial just like for
the Hermitian operators to prove orthogonality or more appropriately,
biorthogonality, in terms of the $\left\{ \left\langle \alpha \right\vert
,\left\vert \beta \right\rangle \right\} $ and $\left\{ \left\langle \beta
\right\vert ,\left\vert \alpha \right\rangle \right\} $ 'dual' eigenspaces.
This seems to suggest the following Hermitian-like relations%
\begin{eqnarray*}
\left\langle \alpha \right\vert \hat{\psi}^{\dagger }\left\vert \beta
\right\rangle &\equiv &\alpha ^{\ast }\left\langle \alpha \right\vert
\left\vert \beta \right\rangle \\
\left\langle \beta \right\vert \hat{\psi}\left\vert \alpha \right\rangle
&\equiv &\alpha \left\langle \beta \right\vert \left\vert \alpha
\right\rangle
\end{eqnarray*}%
where $\left\langle \alpha \right\vert $ and $\left\vert \beta \right\rangle 
$ are the left and right eigenvectors, respectively of $\hat{\psi}^{\dagger
} $, whereas, $\left\langle \beta \right\vert $ and $\left\vert \alpha
\right\rangle $ are the left and right eigenvectors, respectively, of $\hat{%
\psi}$. The above pairing allows us to form dual eigenvectors to simulate
the Hilbert-spaces of Hermitian operators.

We now denote the $\left\langle \alpha \right\vert $ and $\left\vert \beta
\right\rangle $ left and right eigenvectors, respectively of $\hat{\psi}%
^{\dagger }$as making up the $\alpha ^{\ast }$-Hilbert space, and we will
adopt a new consistent labels, $\left\langle \alpha \right\vert
\Longrightarrow \left\langle \mathfrak{p}\right\vert $ and $\left\vert \beta
\right\rangle \Longrightarrow \left\vert \mathfrak{p}\right\rangle $.
Similary, we relabel the $\left\langle \beta \right\vert $ and $\left\vert
\alpha \right\rangle $ left and right eigenvectors, respectively, of $\hat{%
\psi}$ as making up the $\alpha $-Hilbert space, with $\left\langle \beta
\right\vert \Longrightarrow \left\langle \mathfrak{q}\right\vert $ and $%
\left\vert \alpha \right\rangle \Longrightarrow \left\vert \mathfrak{q}%
\right\rangle $. The $\mathfrak{q}$ and $\mathfrak{p}$- eigenspaces
constitute our newly-formed quantum label for Hilbert spaces for $\hat{\psi}$
and $\hat{\psi}^{\dagger }$, respectively.

\subsection{Completeness relations}

In the new dual space representation, Eq. (\ref{one46}) becomes simply a
completeness relation,%
\begin{eqnarray}
\sum\limits_{n}\frac{\left\vert \alpha _{n}\right\rangle \left\langle \beta
_{n}\right\vert }{\left\langle \beta _{n}\right\vert \left\vert \alpha
_{n}\right\rangle } &=&1\Longrightarrow \sum\limits_{\mathfrak{q}}\frac{%
\left\vert \mathfrak{q}\right\rangle \left\langle \mathfrak{q}\right\vert }{%
\left\langle \mathfrak{q}\right\vert \left\vert \mathfrak{q}\right\rangle }=1
\nonumber \\
&\Longrightarrow &\sum\limits_{\mathfrak{q}}\left\vert \mathfrak{q}%
\right\rangle \left\langle \mathfrak{q}\right\vert =1  \label{one47}
\end{eqnarray}%
which yields the completeness relation, with normalized $\left\langle 
\mathfrak{q}\right\vert \left\vert \mathfrak{q}\right\rangle \equiv 1$.
Thus, the the $\left\vert \mathfrak{q}\right\rangle $and $\left\vert 
\mathfrak{p}\right\rangle $eigenstates obey the completenes relations,%
\begin{eqnarray}
\sum\limits_{\mathfrak{q}}\left\vert \mathfrak{q}\right\rangle \left\langle 
\mathfrak{q}\right\vert &=&1  \nonumber \\
\sum\limits_{\mathfrak{p}}\left\vert \mathfrak{p}\right\rangle \left\langle 
\mathfrak{p}\right\vert &=&1  \label{one48}
\end{eqnarray}%
Then it becomes trivial to see the transformation between the dual spaces, $%
\left\vert \mathfrak{q}\right\rangle $and $\left\vert \mathfrak{p}%
\right\rangle $eigenstates,%
\begin{eqnarray}
\left\vert \mathfrak{q}\right\rangle &=&\sum\limits_{\phi }\left\vert 
\mathfrak{p}\right\rangle \left\langle \mathfrak{p}\right\vert \left\vert 
\mathfrak{q}\right\rangle  \label{one49} \\
\left\vert \mathfrak{p}\right\rangle &=&\sum\limits_{\theta }\left\vert 
\mathfrak{q}\right\rangle \left\langle \mathfrak{q}\right\vert \left\vert 
\mathfrak{p}\right\rangle  \label{one50}
\end{eqnarray}%
with transformation function between elements of new dual spaces given by $%
\left\langle \mathfrak{p}\right\vert \left\vert \mathfrak{q}\right\rangle $
and $\left\langle \mathfrak{q}\right\vert \left\vert \mathfrak{p}%
\right\rangle $, respectively. From Eqs. (\ref{one44}) and (\ref{one45}), we
have 
\begin{eqnarray*}
\left\langle \mathfrak{q}_{m}\right\vert \left\vert \mathfrak{q}%
_{n}\right\rangle &=&\delta _{m,n} \\
\left\langle \mathfrak{p}_{m}\right\vert \left\vert \mathfrak{p}%
_{n}\right\rangle &=&\delta _{m,n}
\end{eqnarray*}%
firmly defining the complete and orthogonal dual Hilbert spaces, $\left\{
\left\vert \mathfrak{q}\right\rangle \right\} $ and$\left\{ \left\vert 
\mathfrak{p}\right\rangle \right\} $.

\subsection{Generation of states}

Equation (\ref{one39}), defines the generation of state $\left\vert 
\mathfrak{q}\right\rangle $, 
\begin{equation}
\left\vert \mathfrak{q}\right\rangle =C_{o}\exp \alpha \hat{\psi}^{\dagger
}\left\vert 0\right\rangle  \label{one51}
\end{equation}%
\[
\hat{\psi}^{\dagger }\left\vert \mathfrak{q}\right\rangle =\hat{\psi}%
^{\dagger }\exp \left( q\hat{\psi}^{\dagger }\left\vert 0\right\rangle
\right) =\frac{\partial }{\partial \mathfrak{q}}\mathcal{\ }\left\vert 
\mathfrak{q}\right\rangle 
\]%
Inserting the term $\exp \left\{ -\alpha ^{\ast }\hat{\psi}\right\} $ right
in front of $\left\vert 0\right\rangle $\footnote{%
There is arbitrariness in incorporating $\exp \left\{ \phi \hat{a}\right\} $%
, either positive or negative exponent, operating on vacuum state. To be
symmetric we should use positive exponent, $\exp \left\{ \phi \hat{a}%
\right\} $. For convenience, we want the generation of state unitary, so it
is advisable to use the negative exponent. We will follow this convention is
what follows.}, which has the effect of multiplying by unity, we obtain a
fully symmetric form as 
\begin{equation}
\left\vert \mathfrak{q}\right\rangle =C_{o}\exp \alpha \hat{\psi}^{\dagger
}\exp \left\{ -\alpha ^{\ast }\hat{\psi}\right\} \left\vert \psi
_{0}\right\rangle  \label{one52}
\end{equation}%
To avoid confusion, we set the eigenvalues $\alpha =\mathfrak{q}$ and $%
\alpha ^{\ast }=\mathfrak{p}$. We also set $\left\vert \mathfrak{q}%
\right\rangle =\exp \left( -i\mathfrak{q}\cdot \mathcal{\hat{P}}\right)
\left\vert 0\right\rangle $, i.e., we have,%
\begin{eqnarray}
\hat{\psi}^{\dagger }\left\vert \mathfrak{q}\right\rangle &=&\frac{\partial 
}{\partial \mathfrak{q}}\mathcal{\ }\left\vert \mathfrak{q}\right\rangle =-i%
\mathcal{\hat{P}\ }\left\vert \mathfrak{q}\right\rangle  \label{one53} \\
\mathcal{\hat{P}}\left\vert \mathfrak{p}\right\rangle &=&\ \mathfrak{p}%
\mathcal{\ }\left\vert \mathfrak{p}\right\rangle  \label{one54}
\end{eqnarray}%
To calculate $\left\langle \mathfrak{p}\right\vert \left\vert \mathfrak{q}%
\right\rangle $, we proceed as follows.%
\begin{eqnarray}
\left\langle \mathfrak{p}\right\vert \hat{\psi}^{\dagger }\left\vert 
\mathfrak{q}\right\rangle &=&\left\langle \mathfrak{p}\right\vert \frac{%
\partial }{\partial \mathfrak{q}}\mathcal{\ }\left\vert \mathfrak{q}%
\right\rangle  \label{one55} \\
-i\mathfrak{p}\left\langle \mathfrak{p}\right\vert \left\vert \mathfrak{q}%
\right\rangle &=&\frac{\partial }{\partial \mathfrak{q}}\left\langle 
\mathfrak{p}\right\vert \left\vert \mathfrak{q}\right\rangle  \label{one56}
\\
\frac{\frac{\partial }{\partial \mathfrak{q}}\left\langle \mathfrak{p}%
\right\vert \left\vert \mathfrak{q}\right\rangle }{\left\langle \mathfrak{p}%
\right\vert \left\vert \mathfrak{q}\right\rangle } &=&-i\mathfrak{p}
\label{one57} \\
\frac{\partial }{\partial \mathfrak{q}}\ln \left\langle \mathfrak{p}%
\right\vert \left\vert \mathfrak{q}\right\rangle &=&-i\mathfrak{p}
\label{one58} \\
\left\langle \mathfrak{p}\right\vert \left\vert \mathfrak{q}\right\rangle
&=&\exp \left( -i\mathfrak{q}\cdot \mathfrak{p}\right)  \label{one59}
\end{eqnarray}%
Similarly, we have 
\begin{eqnarray}
\hat{\psi}\left\vert \mathfrak{p}\right\rangle &=&\frac{\partial }{\partial 
\mathfrak{p}}\mathfrak{\ }\left\vert \mathfrak{p}\right\rangle =i\mathfrak{%
\mathcal{\hat{Q}\ }}\left\vert \mathfrak{p}\right\rangle  \label{one60} \\
\mathfrak{\mathcal{\hat{Q}}}\ \left\vert \mathfrak{q}\right\rangle &=&%
\mathfrak{q\ }\left\vert \mathfrak{q}\right\rangle  \label{one61}
\end{eqnarray}%
and%
\begin{eqnarray}
\left\langle \mathfrak{q}\right\vert \hat{\psi}\left\vert \mathfrak{p}%
\right\rangle &=&\left\langle \mathfrak{q}\right\vert \frac{\partial }{%
\partial \mathfrak{p}}\mathfrak{\ }\left\vert \mathfrak{p}\right\rangle 
\nonumber \\
i\mathfrak{q}\left\langle \mathfrak{q}\right\vert \left\vert \mathfrak{p}%
\right\rangle &=&\frac{\partial }{\partial \mathfrak{p}}\mathfrak{\ }%
\left\langle \mathfrak{q}\right\vert \left\vert \mathfrak{p}\right\rangle 
\nonumber \\
\left\langle \mathfrak{q}\right\vert \left\vert \mathfrak{p}\right\rangle
&=&\exp i\mathfrak{p\cdot q}  \label{one62}
\end{eqnarray}%
So far all the above developments holds for fermions and bosons. However,
note that for fermions the eigenvalues corresponding to $\mathfrak{q}$ and $%
\mathfrak{p}$ are elements of the Grassmann algebra. We observe that $\hat{%
\psi}^{\dagger }$dictates the dynamics over $\mathfrak{q}$ space, whereas, $%
\hat{\psi}$ dictates the dynamics over $\mathfrak{p}$ space.

\section{The $\mathfrak{q}$-$\mathfrak{p}$ Representations and Lattice Weyl
Transform}

The mixed $\mathfrak{q}$-$\mathfrak{p}$ representation basically start by
expanding any quantum operator, $\hat{A}$, in terms of mutually unbiased
basis states, namely the eigenvector of annihilation operator, $\hat{\psi}$
or $\mathcal{Q}$, and the eigenvector of creation operator, $\hat{\psi}%
^{\dagger }$or $\mathcal{P}$. We have%
\begin{eqnarray}
\hat{A} &=&\sum\limits_{\mathfrak{p},\mathfrak{q}}\left\vert \mathfrak{q}%
\right\rangle \left\langle \mathfrak{q}\right\vert \hat{A}\left\vert 
\mathfrak{p}\right\rangle \left\langle \mathfrak{p}\right\vert  \nonumber \\
&=&\sum\limits_{\mathfrak{p},\mathfrak{q}}\left\langle \mathfrak{q}%
\right\vert \hat{A}\left\vert \mathfrak{p}\right\rangle \ \left\vert 
\mathfrak{q}\right\rangle \left\langle \mathfrak{p}\right\vert  \label{one68}
\end{eqnarray}

\subsection{The completeness of dual spaces}

The set $\left\{ \left\vert \mathfrak{q}\right\rangle \left\langle \mathfrak{%
p}\right\vert \right\} $ is the basis operators for the mixed $\mathfrak{q}$-%
$\mathfrak{p}$ representation. From the completeness relations of the
unbiased basis states, $\left\{ \left\vert \mathfrak{q}\right\rangle
\right\} $ and $\left\{ \left\vert \mathfrak{p}\right\rangle \right\} $, the
set $\left\{ \left\vert \mathfrak{q}\right\rangle \left\langle \mathfrak{p}%
\right\vert \right\} $ obeys the completeness relation%
\begin{eqnarray}
\sum\limits_{\mathfrak{q},\mathfrak{p}}\left\vert \mathfrak{q}\right\rangle
\left\langle \mathfrak{q}\right\vert \left\vert \mathfrak{p}\right\rangle
\left\langle \mathfrak{p}\right\vert &=&1  \label{one69} \\
\sum\limits_{\mathfrak{q},\mathfrak{p}}\left\langle \mathfrak{q}\right\vert
\left\vert \mathfrak{p}\right\rangle \ \left\vert \mathfrak{q}\right\rangle
\left\langle \mathfrak{p}\right\vert &=&1  \label{one70}
\end{eqnarray}%
Substituting the expression for $\left\langle \mathfrak{q}\right\vert
\left\vert \mathfrak{p}\right\rangle $, 
\begin{eqnarray}
\left\langle \mathfrak{q}\right\vert \left\vert \mathfrak{p}\right\rangle
&=&\exp \left( i\mathfrak{p}\cdot \mathfrak{q}\right)  \label{one71} \\
\left\langle \mathfrak{p}\right\vert \left\vert \mathfrak{q}\right\rangle
&=&\exp \left( -i\mathfrak{p}\cdot \mathfrak{q}\right)  \label{one72}
\end{eqnarray}%
we obtained, for the completeness relation, 
\begin{equation}
C_{0}\sum\limits_{\mathfrak{q},\mathfrak{p}}\exp \left( i\mathfrak{p}\cdot 
\mathfrak{q}\right) \left\vert \mathfrak{q}\right\rangle \left\langle 
\mathfrak{p}\right\vert =1  \label{one73}
\end{equation}%
where $C_{o}$can be choosen as 
\[
C_{o}=\left( N\right) ^{-\frac{1}{2}} 
\]%
Equation (\ref{one73}) can be rewritten as%
\begin{equation}
\left( N\right) ^{-\frac{1}{2}}\sum\limits_{\mathfrak{q},\mathfrak{p}}\frac{%
\left\vert \mathfrak{q}\right\rangle \left\langle \mathfrak{p}\right\vert }{%
\left\langle \mathfrak{p}\right\vert \left\vert \mathfrak{q}\right\rangle }=1
\label{one74}
\end{equation}%
and similarly,%
\[
\left( N\right) ^{-\frac{1}{2}}\sum\limits_{\mathfrak{q},\mathfrak{p}}\frac{%
\left\vert \mathfrak{p}\right\rangle \left\langle \mathfrak{q}\right\vert }{%
\left\langle \mathfrak{q}\right\vert \left\vert \mathfrak{p}\right\rangle }%
=1 
\]%
Here we use the transformation identities in the mixed $\mathfrak{q}$-$%
\mathfrak{p}$ representation, 
\begin{eqnarray}
\left\vert \mathfrak{p}\right\rangle &=&\sum\limits_{\mathfrak{q}%
}\left\langle \mathfrak{q}\right\vert \left\vert \mathfrak{p}\right\rangle \
\left\vert \mathfrak{q}\right\rangle  \label{one75} \\
\left\langle \mathfrak{p}\right\vert &=&\sum\limits_{\mathfrak{q}%
}\left\langle \mathfrak{p}\right\vert \left\vert \mathfrak{q}\right\rangle \
\left\langle \mathfrak{q}\right\vert  \label{one76} \\
\left\vert \mathfrak{q}\right\rangle &=&\sum\limits_{\mathfrak{p}%
}\left\langle \mathfrak{p}\right\vert \left\vert \mathfrak{q}\right\rangle \
\left\vert \mathfrak{p}\right\rangle  \label{one77} \\
\left\langle \mathfrak{q}\right\vert &=&\sum\limits_{\mathfrak{p}%
}\left\langle \mathfrak{q}\right\vert \left\vert \mathfrak{p}\right\rangle \
\left\langle \mathfrak{p}\right\vert  \label{one78}
\end{eqnarray}%
with transformation functions given by Eqs. (\ref{one71})-(\ref{one72}).

\subsection{The expansion of any operators in dual space}

Any operator, $\hat{A}$, can be expressed in terms of the unbiased
eigenvector spaces, $\left\vert \mathfrak{q}^{\prime }\right\rangle $and $%
\left\vert \mathfrak{p}^{\prime \prime }\right\rangle $, respectively, in a
mixed representation by Eq. (\ref{one68}), which we rewrite as,

\begin{eqnarray}
\hat{A} &=&\sum\limits_{\mathfrak{p}^{\prime \prime },\mathfrak{q}^{\prime
}}\left\vert \mathfrak{p}^{\prime \prime }\right\rangle \left\langle 
\mathfrak{p}^{\prime \prime }\right\vert A\left\vert \mathfrak{q}^{\prime
}\right\rangle \left\langle \mathfrak{q}^{\prime }\right\vert  \nonumber \\
&=&\sum\limits_{\mathfrak{p}^{\prime \prime },\mathfrak{q}^{\prime
}}\left\langle \mathfrak{p}^{\prime \prime }\right\vert A\left\vert 
\mathfrak{q}^{\prime }\right\rangle \ \left\vert \mathfrak{p}^{\prime \prime
}\right\rangle \left\langle \mathfrak{q}^{\prime }\right\vert  \label{one79}
\end{eqnarray}%
We wish to express $\left\langle \mathfrak{p}^{\prime \prime }\right\vert
A\left\vert \mathfrak{q}^{\prime }\right\rangle $ and $\ \left\vert 
\mathfrak{p}^{\prime \prime }\right\rangle \left\langle \mathfrak{q}^{\prime
}\right\vert $ in terms of the $\left\vert \mathfrak{q}\right\rangle $%
-eigenstate matrix elements and $\left\vert \mathfrak{p}\right\rangle $%
-space projectors, respectively. Using, Eqs. (\ref{one75})-(\ref{one78}), we
write%
\begin{eqnarray}
\left\langle \mathfrak{p}^{\prime \prime }\right\vert A\left\vert \mathfrak{q%
}^{\prime }\right\rangle &=&\frac{1}{\sqrt{N}}\sum\limits_{\mathfrak{q}%
^{\prime \prime }}e^{-i\mathfrak{p}^{\prime \prime }\cdot \mathfrak{q}%
^{\prime \prime }}\left\langle \mathfrak{q}^{\prime \prime }\right\vert
A\left\vert \mathfrak{q}^{\prime }\right\rangle  \nonumber \\
\left\vert \mathfrak{p}^{\prime \prime }\right\rangle \left\langle \mathfrak{%
q}^{\prime }\right\vert &=&\frac{1}{\sqrt{N}}\sum\limits_{\mathfrak{p}%
^{\prime }}e^{i\mathfrak{p}^{\prime }\cdot \mathfrak{q}^{\prime }}\left\vert 
\mathfrak{p}^{\prime \prime }\right\rangle \left\langle \mathfrak{p}^{\prime
}\right\vert  \label{one80}
\end{eqnarray}%
with completeness relation, using the transformation function characteristic
of dual spaces,%
\[
\frac{1}{\sqrt{N}}\sum\limits_{\mathfrak{q},\mathfrak{p}}\exp \left( i%
\mathfrak{p}\cdot \mathfrak{q}\right) \left\vert \mathfrak{q}\right\rangle
\left\langle \mathfrak{p}\right\vert =1=\sum\limits_{\mathfrak{p}}\left\vert 
\mathfrak{p}\right\rangle \left\langle \mathfrak{p}\right\vert 
\]%
Introducing the notation in Eq. (\ref{one80}),%
\begin{eqnarray*}
\mathfrak{p}^{\prime } &=&\mathfrak{p}+\mathfrak{u}\text{, \ \ \ \ \ \ \ }%
\mathfrak{q}^{\prime }=\mathfrak{q}+\mathfrak{v}, \\
\mathfrak{p}^{\prime \prime } &=&\mathfrak{p}-\mathfrak{u}\text{, \ \ \ \ \
\ \ }\mathfrak{q}^{\prime \prime }=\mathfrak{q}-\mathfrak{v}\text{.}
\end{eqnarray*}%
Then, upon substituting in Eq. (\ref{one79}), we end up with%
\begin{eqnarray}
A &=&\sum\limits_{\mathfrak{p}^{\prime \prime },\mathfrak{q}^{\prime
}}\left\vert \mathfrak{p}^{\prime \prime }\right\rangle \left\langle 
\mathfrak{p}^{\prime \prime }\right\vert A\left\vert \mathfrak{q}^{\prime
}\right\rangle \left\langle \mathfrak{q}^{\prime }\right\vert  \nonumber \\
&=&\frac{1}{N}\sum\limits_{\mathfrak{p},\mathfrak{q},\mathfrak{u},\mathfrak{v%
}}e^{i2\left( \mathfrak{p}\cdot \mathfrak{v}+\mathfrak{u}\cdot \mathfrak{q}%
\right) }\left\langle \mathfrak{q}-\mathfrak{v}\right\vert A\left\vert 
\mathfrak{q}+\mathfrak{v}\right\rangle \ \left\vert \mathfrak{p}-\mathfrak{u}%
\right\rangle \left\langle \mathfrak{p}+\mathfrak{u}\right\vert
\label{op_in_pq_eigenf}
\end{eqnarray}

\subsection{Mixed space operator basis, $\hat{\Delta}\left( \mathfrak{p},%
\mathfrak{q}\right) $}

We write the last result as an expansion in terms of \textit{mixed}-\textit{%
phase point projector}, $\hat{\Delta}\left( \mathfrak{p},\mathfrak{q}\right) 
$, defined as the\textit{\ Weyl transform} of a projector, by%
\begin{eqnarray}
\hat{\Delta}\left( \mathfrak{p},\mathfrak{q}\right) &=&\sum\limits_{%
\mathfrak{u}}e^{-i2\mathfrak{u}\cdot \mathfrak{q}}\left\vert \mathfrak{p}+%
\mathfrak{u}\right\rangle \left\langle \mathfrak{p}-\mathfrak{u}\right\vert 
\nonumber \\
&=&\sum\limits_{\mathfrak{u}}e^{-i2\mathfrak{u}\cdot \mathfrak{q}%
}e^{2iQ\cdot u}\left\vert \mathfrak{p}-\mathfrak{u}\right\rangle
\left\langle \mathfrak{p}-\mathfrak{u}\right\vert  \nonumber \\
&=&\sum\limits_{\mathfrak{u}}e^{-i2\mathfrak{u\cdot q}}e^{2i\mathcal{Q}\cdot 
\mathfrak{u}}\sum\limits_{\mathfrak{v}}e^{2i\left( \mathfrak{p-u}-\mathcal{P}%
\right) \cdot \mathfrak{v}}\left\vert \mathfrak{p}_{0}\right\rangle
\left\langle \mathfrak{p}_{0}\right\vert  \nonumber \\
&=&\sum\limits_{\mathfrak{u,v}}e^{2i\left( \mathcal{Q-}\mathfrak{q}\right)
\cdot \mathfrak{u}}e^{-2i\left( \mathcal{P-}\mathfrak{p}\right) \cdot 
\mathfrak{v}}e^{-2i\mathfrak{u}\cdot \mathfrak{v}}\left\vert \mathfrak{p}%
_{0}\right\rangle \left\langle \mathfrak{p}_{0}\right\vert  \nonumber \\
&=&\sum\limits_{\mathfrak{u,v}}e^{2i\left( \mathfrak{p}\cdot \mathfrak{%
v-q\cdot u}\right) }e^{-2i\left( \mathcal{P}\cdot \mathfrak{v-}\mathcal{%
Q\cdot }\mathfrak{u}\right) }\sum\limits_{\mathfrak{p}_{0}}\left\vert 
\mathfrak{p}_{0}\right\rangle \left\langle \mathfrak{p}_{0}\right\vert
\label{one81}
\end{eqnarray}%
and the coefficient of expansion, the so-called \textit{Weyl transform} of
matrix element of operator, $A\left( \mathfrak{p},\mathfrak{q}\right) $,
defined by%
\begin{equation}
A\left( \mathfrak{p},\mathfrak{q}\right) =\sum\limits_{\mathfrak{v}}e^{i2%
\mathfrak{p}\cdot \mathfrak{v}}\left\langle \mathfrak{q}-\mathfrak{v}%
\right\vert A\left\vert \mathfrak{q}+\mathfrak{v}\right\rangle \text{.}
\label{lwt}
\end{equation}%
Clearly, for a density matrix operator $\hat{\rho}$, the Weyl transform
obeys,%
\[
\sum\limits_{\mathfrak{p},\mathfrak{q}}\rho \left( \mathfrak{p},\mathfrak{q}%
\right) =1 
\]%
If one accounts for other extra discrete quantum labels like spin and
energy-band indices, we can incorporate this in the summation in a form of a
trace.

Thus, we eventually have any operator expanded in terms of mixed space
operator basis, $\hat{\Delta}\left( \mathfrak{p},\mathfrak{q}\right) $,%
\begin{eqnarray}
\hat{A} &=&\sum\limits_{\mathfrak{p},\mathfrak{q}}A\left( \mathfrak{p},%
\mathfrak{q}\right) \ \hat{\Delta}\left( \mathfrak{p},\mathfrak{q}\right) 
\nonumber \\
&=&\sum\limits_{\mathfrak{u,v}}\left( \sum\limits_{\mathfrak{p},\mathfrak{q}%
}A\left( \mathfrak{p},\mathfrak{q}\right) e^{2i\left( \mathfrak{p}\cdot 
\mathfrak{v-q\cdot u}\right) }\right) e^{-2i\left( \mathcal{P}\cdot 
\mathfrak{v-}\mathcal{Q\cdot }\mathfrak{u}\right) }\left\vert \mathfrak{p}%
_{0}\right\rangle \left\langle \mathfrak{p}_{0}\right\vert  \label{one82}
\end{eqnarray}%
We have%
\begin{eqnarray*}
A\left( \mathfrak{p},\mathfrak{q}\right) &=&Tr\left( \hat{A}\ \hat{\Delta}%
\left( \mathfrak{p},\mathfrak{q}\right) \right) \\
&=&\ \sum\limits_{\mathfrak{u,v}}e^{2i\left( \mathfrak{p}\cdot \mathfrak{%
v-q\cdot u}\right) }Tr\left( \hat{A}e^{-2i\left( \mathcal{P}\cdot \mathfrak{%
v-}\mathcal{Q\cdot }\mathfrak{u}\right) }\left\vert \mathfrak{p}%
_{0}\right\rangle \left\langle \mathfrak{p}_{0}\right\vert \right) \\
&=&\sum\limits_{\mathfrak{u,v}}e^{2i\left( \mathfrak{p}\cdot \mathfrak{%
v-q\cdot u}\right) }A\left( \mathfrak{u,v}\right)
\end{eqnarray*}%
where $A\left( \mathfrak{u,v}\right) $ is the characteristic function of $%
A\left( \mathfrak{p},\mathfrak{q}\right) $ distribution. Upon similar
procedure based on Eq. (\ref{one79}), an equivalent expression can be obtain
for $A\left( \mathfrak{p},\mathfrak{q}\right) $ and $\hat{\Delta}\left( 
\mathfrak{p},\mathfrak{q}\right) $, namely,%
\begin{eqnarray}
A\left( \mathfrak{p},\mathfrak{q}\right) &=&\sum\limits_{\mathfrak{u}}e^{i2%
\mathfrak{u}\cdot \mathfrak{q}}\left\langle \mathfrak{p}+\mathfrak{u}%
\right\vert \hat{A}\left\vert \mathfrak{p}-\mathfrak{u}\right\rangle
\label{one83} \\
\hat{\Delta}\left( \mathfrak{p},\mathfrak{q}\right) &=&\sum\limits_{%
\mathfrak{v}}e^{i2\mathfrak{p}\cdot \mathfrak{v}}\left\vert \mathfrak{q}+%
\mathfrak{v}\right\rangle \left\langle \mathfrak{q}-\mathfrak{v}\right\vert
\label{one84}
\end{eqnarray}

We can combine the exponential operators to obtain%
\begin{eqnarray*}
&&\left( N\right) ^{-1}\sum\limits_{\bar{v},\bar{u}}e^{2i\mathfrak{p}\cdot 
\mathfrak{v}}\exp \left[ -2i\mathcal{P}\cdot \mathfrak{v}\right] \exp \left[
2i\left( \mathfrak{q}-\mathfrak{v}-\mathcal{Q}\right) \cdot \mathfrak{\bar{u}%
}\right] \sum\limits_{\mathfrak{q}_{0}}\left\vert \mathfrak{q}_{0},\lambda
\right\rangle \left\langle \mathfrak{q}_{0},\lambda ^{\prime }\right\vert \\
&=&\left( N\right) ^{-1}\sum\limits_{\bar{v},\bar{u}}\exp -2i\left[ \left( 
\mathcal{P}-\mathfrak{p}\right) \cdot \mathfrak{v}+\left( \mathcal{Q}-%
\mathfrak{q}\right) \cdot \mathfrak{u}\right] \Omega _{\lambda \lambda
^{\prime }}
\end{eqnarray*}%
where, 
\[
\Omega _{\lambda \lambda ^{\prime }}=\sum\limits_{\mathfrak{q}%
_{0}}\left\vert \mathfrak{q}_{0},\lambda \right\rangle \left\langle 
\mathfrak{q}_{0},\lambda ^{\prime }\right\vert =\sum\limits_{\mathfrak{p}%
_{0}}\left\vert \mathfrak{p}_{0},\lambda \right\rangle \left\langle 
\mathfrak{p}_{0},\lambda ^{\prime }\right\vert 
\]%
\begin{eqnarray*}
A\left( \mathfrak{p},\mathfrak{q}\right) &=&Tr\left( \hat{A}\hat{\Delta}%
\right) \\
&=&\left( N\right) ^{-1}\left( 
\begin{array}{c}
\sum\limits_{\bar{v},\bar{u}}\exp 2i\left[ \mathfrak{p}\cdot \mathfrak{v}-%
\mathfrak{q}\cdot \mathfrak{u}\right] \  \\ 
\times Tr\left\{ \hat{A}\exp \left\{ -2i\left[ \mathcal{P}\cdot \mathfrak{v}-%
\mathcal{Q}\cdot \mathfrak{u}\right] \right\} \Omega _{\lambda \lambda
^{\prime }}\right\}%
\end{array}%
\right)
\end{eqnarray*}%
Therefore, the charactetic distribution for $A\left( p,q\right) $is
identically, 
\[
A_{\lambda \lambda ^{\prime }}\left( \mathfrak{u},\mathfrak{v}\right)
=Tr\left\{ \hat{A}\exp \left\{ -2i\left[ \mathcal{P}\cdot \mathfrak{v}-%
\mathcal{Q}\cdot \mathfrak{u}\right] \right\} \Omega _{\lambda \lambda
^{\prime }}\right\} 
\]%
as before. By using the characteristic distribution function for $A_{\lambda
\lambda ^{\prime }}\left( \mathfrak{p},\mathfrak{q}\right) $ 
\begin{equation}
A_{\lambda \lambda ^{\prime }}\left( \mathfrak{u},\mathfrak{v}\right)
=\left( \frac{1}{N}\right) ^{\frac{1}{2}}\sum\limits_{p,q}A_{\lambda \lambda
^{\prime }}\left( \mathfrak{p},\mathfrak{q}\right) \exp 2i\left[ \mathfrak{p}%
\cdot \mathfrak{v}-\mathfrak{q}\cdot \mathfrak{u}\right]  \label{one94}
\end{equation}%
with inverse%
\[
A_{\lambda \lambda ^{\prime }}\left( \mathfrak{p},\mathfrak{q}\right)
=\left( \frac{1}{N}\right) ^{\frac{1}{2}}\sum\limits_{u,v}A_{\lambda \lambda
^{\prime }}\left( \mathfrak{u},\mathfrak{v}\right) \exp \left\{ -2i\left[ 
\mathfrak{p}\cdot \mathfrak{v}-\mathfrak{q}\cdot \mathfrak{u}\right]
\right\} 
\]%
Then we can write Eq. (\ref{one82}) simply like a Fourier transform (caveat:
Fourier transform to\textit{\ operator space}) of the characteristic
function of the lattice Weyl transform of the operator $\hat{A}$, 
\begin{equation}
\hat{A}=\sum\limits_{\bar{u},\bar{v},\lambda ,\lambda ^{\prime }}A_{\lambda
\lambda ^{\prime }}\left( \mathfrak{u},\mathfrak{v}\right) \exp \left\{ -2i%
\left[ \mathcal{P}\cdot \mathfrak{v}-\mathcal{Q}\cdot \mathfrak{u}\right]
\right\} \Omega _{\lambda ^{\prime }\lambda }  \label{one95}
\end{equation}%
where the inverse can be written as%
\[
A_{\lambda \lambda ^{\prime }}\left( \mathfrak{u},\mathfrak{v}\right)
=Tr\left\{ \hat{A}\exp \left\{ -2i\left[ \mathcal{P}\cdot \mathfrak{v}-%
\mathcal{Q}\cdot \mathfrak{u}\right] \right\} \right\} \Omega _{\lambda
\lambda ^{\prime }} 
\]%
In continuum approximation, we have,%
\begin{eqnarray}
\hat{\Delta}\left( \mathfrak{p,q}\right) &=&\left( 2\pi \right) ^{-1}\int d%
\mathfrak{u}\wedge d\mathfrak{v}\ e^{2i\left[ \left( \ \left( \mathfrak{p}-%
\mathcal{P}\right) .v-\mathfrak{q}-\mathcal{Q}\right) .\mathfrak{u}\ \right]
}  \nonumber \\
&=&\left( 2\pi \right) ^{-1}\int d\mathfrak{u}\wedge d\mathfrak{v}\ e^{2i%
\left[ \mathfrak{p}\cdot \mathfrak{v}-\mathfrak{q}\cdot \mathfrak{u}\right]
}e^{\left( -2i\right) \left( \mathcal{P}\cdot \mathfrak{v}-\mathcal{Q}\cdot 
\mathfrak{u}\right) },  \label{one96}
\end{eqnarray}

\subsection{On the counting of states: Coherent state formulation}

The coherent states formulation is a special case of mixed representation in
QFT. The measure of counting of states in Eq. (\ref{one96}) can be shown to
be related to original annihilation and creation operator in the case of
harmonic oscillator. We have 
\begin{eqnarray*}
\left( 2\pi \right) ^{-1}\int d\mathfrak{u}\wedge d\mathfrak{v} &\mathfrak{%
\Longrightarrow }&\left( 2\pi \right) ^{-1}\int -id\mathfrak{u}\wedge d%
\mathfrak{v} \\
&=&\left( 2\pi \right) ^{-1}\int \frac{1}{2}d\left( \mathfrak{\bar{q}-i\bar{p%
}}\right) \wedge d\left( \mathfrak{\bar{q}+i\bar{p}}\right) \\
&=&\left( 2\pi \right) ^{-1}\int \left( d\mathfrak{\bar{q}\wedge }d\mathfrak{%
\bar{p}}\right) \\
&=&\frac{1}{\pi }\int \left( d\func{Re}\alpha \wedge d\func{Im}\alpha \right)
\end{eqnarray*}%
since $d\func{Re}\alpha =\frac{d\mathfrak{\bar{q}}}{\sqrt{2}}$ and $d\func{Im%
}\alpha =\frac{d\mathfrak{\bar{p}}}{\sqrt{2}}$.

\section{Characteristic Distribution of Lattice Weyl Transform}

Therefore, we have the identity for the characteristic function $A_{\lambda
\lambda ^{\prime }}\left( \mathfrak{u},\mathfrak{v}\right) $%
\begin{equation}
A_{\lambda \lambda ^{\prime }}\left( \mathfrak{u},\mathfrak{v}\right)
=Tr\left\{ \hat{A}\exp \left\{ -2i\left[ \mathcal{P}\cdot \mathfrak{v}-%
\mathcal{Q}\cdot \mathfrak{u}\right] \right\} \right\} \Omega _{\lambda
\lambda ^{\prime }}  \label{one97}
\end{equation}%
The characteristic function exist for all function of canonical quantum
operators, either Hermitian or non-Hermitian, spinor (fermions) or boson
operators \footnote{%
For creation and annihilation operators in many-body quantum physics, the
proof relies on the use of normal or anti-normal ordering of canonical
operators, which can then be treated like $%
%TCIMACRO{\U{2102} }%
%BeginExpansion
\mathbb{C}
%EndExpansion
$-numbers in expansion of exponentials. The exponential in Eq. (\ref{one97})
is sometimes referred to as the generalized Pauli-spin operator.}.

\subsection{Implications on coherent states (CS) formulation}

In what follows, we will drop the discrete indices $\lambda $ and $\lambda
^{\prime }$ to make contact with CS formulation of quantum physics. In
general, we can have different expression for the characteristic function
depending on the use of, what is often referred to in corresponding CS
formalism as the normal and anti-normal expessions,%
\begin{eqnarray*}
\exp \left\{ -2i\left[ \mathcal{P}\cdot \mathfrak{v}-\mathcal{Q}\cdot 
\mathfrak{u}\right] \right\} &=&\exp \left\{ i\mathfrak{u}\cdot \mathfrak{v}%
\right\} \exp \left\{ -2i\mathfrak{v}\cdot \mathcal{P}\right\} \exp \left\{
2i\mathfrak{u}\cdot \mathcal{Q}\right\} \\
&=&\exp \left\{ -i\mathfrak{u}\cdot \mathfrak{v}\right\} \exp \left\{ 2i%
\mathfrak{u}\cdot \mathcal{Q}\right\} \exp \left\{ -2i\mathfrak{v}\cdot 
\mathcal{P}\right\}
\end{eqnarray*}%
so that 
\begin{eqnarray}
\exp \left\{ -2i\mathfrak{v}\cdot \mathcal{P}\right\} \exp \left\{ 2i%
\mathfrak{u}\cdot \mathcal{Q}\right\} &=&\exp \left\{ -i\mathfrak{u}\cdot 
\mathfrak{v}\right\} \exp \left\{ -2i\left[ \mathcal{P}\cdot \mathfrak{v}-%
\mathcal{Q}\cdot \mathfrak{u}\right] \right\}  \label{one98} \\
\exp \left\{ 2i\mathfrak{u}\cdot \mathcal{Q}\right\} \exp \left\{ -2i%
\mathfrak{v}\cdot \mathcal{P}\right\} &=&\exp \left\{ i\mathfrak{u}\cdot 
\mathfrak{v}\right\} \exp \left\{ -2i\left[ \mathcal{P}\cdot \mathfrak{v}-%
\mathcal{Q}\cdot \mathfrak{u}\right] \right\}  \label{one99}
\end{eqnarray}%
yielding the following differrent expressions for $A_{\lambda \lambda
^{\prime }}\left( u,v\right) $, namely, 
\begin{equation}
A_{\lambda \lambda ^{\prime }}^{w}\left( u,v\right) =Tr\left( \hat{A}\exp
\left( -2i\right) \left( \mathcal{P}.\mathfrak{v-}\mathcal{Q}.\mathfrak{u}\
\right) \right)  \label{one100}
\end{equation}%
which is the characteristic function for the Wigner distribution function.
We also have the so-called normal characteristic distribution function, 
\begin{eqnarray}
A_{\lambda \lambda ^{\prime }}^{n}\left( u,v\right) &=&Tr\left[ \ \hat{A}%
\exp \left\{ -2i\mathfrak{v}\cdot \mathcal{P}\right\} \exp \left\{ 2i%
\mathfrak{u}\cdot \mathcal{Q}\right\} \right] ,  \nonumber \\
&=&\exp \left\{ -i\mathfrak{u}\cdot \mathfrak{v}\right\} Tr\left( \hat{A}%
\exp \left( -2i\right) \left( \mathcal{P}.\mathfrak{v-}\mathcal{Q}.\mathfrak{%
u}\ \right) \right)  \label{one101}
\end{eqnarray}%
and the anti-normal characteristic distribution function given by%
\begin{eqnarray}
A_{\lambda \lambda ^{\prime }}^{a}\left( u,v\right) &=&Tr\ \left\{ \hat{A}\
\exp \left\{ 2i\mathfrak{u}\cdot \mathcal{Q}\right\} \exp \left\{ -2i%
\mathfrak{v}\cdot \mathcal{P}\right\} \right\} ,  \nonumber \\
&=&\exp \left\{ i\mathfrak{u}\cdot \mathfrak{v}\right\} Tr\left( \hat{A}\exp
\left( -2i\right) \left( \mathcal{P}.\mathfrak{v-}\mathcal{Q}.\mathfrak{u}%
\right) \right)  \label{one102}
\end{eqnarray}%
Although, Eqs. (\ref{one101}) and (\ref{one102}) only amounts to difference
in the phase factors in the canonical position-momentum $q$-$p$ ordinary
mixed space representation, similar quantities in non-Hermitian dual spaces
gives a real exponents giving very different distributions often referred \
to as smooth-out distributions. Examining Eqs. (\ref{one98}) and (\ref{one99}%
), and the fact that in non-Hermitian mixed representation of the coherent
state,%
\begin{eqnarray}
\exp \left\{ -i\mathfrak{u}\cdot \mathfrak{v}\right\} &\equiv &\bar{\alpha}%
^{\ast }\bar{\alpha}=\frac{1}{2}\left( \bar{q}-i\bar{p}\right) \left( \bar{q}%
+i\bar{p}\right)  \nonumber \\
&=&\frac{1}{2}\left( \func{Re}\bar{\alpha}^{2}+\func{Im}\bar{\alpha}%
^{2}\right) \text{,}  \label{one103} \\
\exp \left\{ i\mathfrak{u}\cdot \mathfrak{v}\right\} &\equiv &-\bar{\alpha}%
^{\ast }\bar{\alpha}=-\frac{1}{2}\left( \bar{q}-i\bar{p}\right) \left( \bar{q%
}+i\bar{p}\right)  \nonumber \\
&=&-\frac{1}{2}\left( \func{Re}\bar{\alpha}^{2}+\func{Im}\bar{\alpha}%
^{2}\right) \text{,}  \label{one104}
\end{eqnarray}%
in original notation is a real quantity that resembles a Guassian function,
Eqs. (\ref{one99}) clearly represent some smoothing of the Wigner
distribution characteristic function and hence the Wigner distribution
itself. Indeed, in Eqs. (\ref{one101}) and (\ref{one102}) no informations
are lost.

Indeed, more \textit{general} phase-space distribution functions, $f^{\left(
g\right) }\left( \mathfrak{p,q},t\right) $, can be obtained from the
expression%
\begin{eqnarray}
f^{\left( g\right) }\left( \mathfrak{p,q},t\right) &=&\left( \frac{1}{N}%
\right) ^{\frac{1}{2}}\sum\limits_{u,v}A_{\lambda \lambda ^{\prime }}\left( 
\mathfrak{u},\mathfrak{v}\right) \exp \left\{ -2i\left[ \mathfrak{p}\cdot 
\mathfrak{v}-\mathfrak{q}\cdot \mathfrak{u}\right] \right\} g\left( 
\mathfrak{u,v}\right)  \nonumber \\
&=&\frac{1}{2\pi }\int dudve^{-i\left( \left[ \mathfrak{p}\cdot \mathfrak{v}-%
\mathfrak{q}\cdot \mathfrak{u}\right] \right) }\left[ C^{\left( w\right)
}\left( \mathfrak{u,v},t\right) \ g\left( \mathfrak{u,v}\right) \right] ,
\label{one105}
\end{eqnarray}%
where $g\left( \mathfrak{u,v}\right) $is some choosen smoothing function.

\subsection{$P$- and $Q$-distribution function: Husimi distribution and
smoothing}

Generally all distribution function will become meaningful under the
integral sign, thus these have the properties of generalized distribution
functions. The distribution function $f^{a}\left( \mathfrak{p,q},t\right) $
that one obtain from Eq. (\ref{one103}) is known as the $P$-function and
that obtain from Eq. (\ref{one104}) is also known as the $Q$-function or the
Husimi distribution in quantum optics. A detailed discussion of these two
distribution is given by the author's book on the topic of coherent state
formulation, and will not be repeated here.

The generalized spin operator algebra of $\exp \left\{ -2i\left( q^{\prime
}\cdot P-p^{\prime }\cdot Q\right) \right\} $ as well as its relevance to
the physics of two-state systems, spin systems, quantum computing,
entanglements \cite{bk:ref20} and teleportation, are discussed in the
author's book \cite{bk:ref8}

\section{A Third Quantization Scheme}

To investigate the quantum dynamics obeyed by the $Q$-distribution in $%
\left( \mathfrak{p},\mathfrak{q}\right) $-space of QFT, we employ a third
quantization scheme. Based on Eq. (\ref{one68}), we derive a third
quantization wavefunction $\psi \left( \mathfrak{q},t\right) $ where $%
\mathfrak{q}$ is the "position" eigenvalue (i.e., eigenvalue of \textit{%
second quantized} annihilation operator) and $t$ denotes the time variable.

To construct the third quantization annihilation and creation field operator 
$\check{\psi}\left( \mathfrak{q},t\right) $ and $\check{\psi}^{\dagger
}\left( \mathfrak{q},t\right) $, we decompose a general field operators $%
\tilde{\Psi}\left( t\right) $ in terms of the eigenfunctions, $\left\vert 
\mathfrak{q}\right\rangle $, of the annihilation operator. We have, i.e.,%
\begin{eqnarray}
\tilde{\Psi}\left( t\right) &=&\dsum\limits_{\mathfrak{q}}\check{\psi}\left( 
\mathfrak{q},t\right) \left\vert \mathfrak{q}\right\rangle  \label{q-decomp}
\\
\tilde{\Psi}^{\dagger }\left( t\right) &=&\dsum\limits_{\mathfrak{p}}\check{%
\psi}^{\dagger }\left( \mathfrak{p},t\right) \left\langle \mathfrak{p}%
\right\vert  \label{p-decomp}
\end{eqnarray}%
where the presence of $x$ in the argument of the general field operators, $%
\tilde{\Psi}\left( t\right) ,$ and $\left\vert \mathfrak{q}\right\rangle $
is just suppressed. In Eq. (\ref{p-decomp}), we used $\left\langle \mathfrak{%
p}\right\vert $ since this is associated with the left eigenfunction of the 
\textit{second} quantization creation, $\hat{\psi}^{\dagger }$. However, we
can transform $\left\langle \mathfrak{p}\right\vert $ to $\left\langle 
\mathfrak{q}\right\vert $. Using Eq. (\ref{one76}), we have, 
\[
\tilde{\Psi}^{\dagger }\left( t\right) =\dsum\limits_{\mathfrak{p}}\check{%
\psi}^{\dagger }\left( \mathfrak{p},t\right) \sum\limits_{\mathfrak{q}%
}\left\langle \mathfrak{p}\right\vert \left\vert \mathfrak{q}\right\rangle \
\left\langle \mathfrak{q}\right\vert 
\]%
Upon interchanging the order of summation, we have 
\begin{eqnarray}
&=&\sum\limits_{\mathfrak{q}}\left\{ \dsum\limits_{\mathfrak{p}}\left\langle 
\mathfrak{p}\right\vert \left\vert \mathfrak{q}\right\rangle \check{\psi}%
^{\dagger }\left( \mathfrak{p},t\right) \right\} \ \left\langle \mathfrak{q}%
\right\vert  \nonumber \\
&=&\sum\limits_{\mathfrak{q}}\left\{ \dsum\limits_{\mathfrak{p}}\exp \left(
-i\mathfrak{p}\cdot \mathfrak{q}\right) \check{\psi}^{\dagger }\left( 
\mathfrak{p},t\right) \right\} \ \left\langle \mathfrak{q}\right\vert 
\nonumber \\
&=&\sum\limits_{\mathfrak{q}}\check{\psi}^{\dagger }\left( \mathfrak{q}%
,t\right) \ \left\langle \mathfrak{q}\right\vert  \label{p-decomp-2}
\end{eqnarray}%
Thus, $\check{\psi}^{\dagger }\left( \mathfrak{q},t\right) $ and $\check{\psi%
}\left( \mathfrak{q},t\right) $, annihilate and create second quantized
fields indicated by $\left\vert \mathfrak{q}\right\rangle $ and its dual $%
\left\langle \mathfrak{q}\right\vert $. This was interpreted as equivalent
to second-quantized coupling constants in Ref. \cite{goodstrom} in their
application of third quantization to cosmology. These third quantized field
operators, $\check{\psi}\left( \mathfrak{q},t\right) $ and $\check{\psi}%
^{\dagger }\left( \mathfrak{q},t\right) $, satisfy the equal-time
commutation relation for bosons and anticommutation for fermions, namely,%
\[
\left[ \check{\psi}\left( \mathfrak{q}\right) ,\check{\psi}^{\dagger }\left( 
\mathfrak{q}^{\prime }\right) \right] _{\pm }=\delta _{\mathfrak{q,q}%
^{\prime }} 
\]%
\[
\left[ \check{\psi}\left( \mathfrak{q}\right) ,\check{\psi}\left( \mathfrak{q%
}^{\prime }\right) \right] _{\pm }=0\text{,} 
\]%
\[
\left[ \check{\psi}^{\dagger }\left( \mathfrak{q}\right) ,\check{\psi}%
^{\dagger }\left( \mathfrak{q}^{\prime }\right) \right] _{\pm }=0 
\]%
where the $+$ subscript is for fermions and the $-$ is bosons.

\section{Transport Dynamics of the $Q$-Distribution of QFT}

Following the usual prescription of second quantization procedure, i.e., by
writing a \textit{single} particle operator in terms of field operators, we
have 
\begin{equation}
\hat{A}^{\left( 1\right) }=\dint \hat{\Psi}^{\dagger }\left( x\right)
A^{\left( 1\right) }\hat{\Psi}\left( x\right) dx  \label{single_op}
\end{equation}%
Likewise, in the third quantization procedure, we can write in the same way,
where $A^{\left( 1\right) }$ in $\hat{\Psi}^{\dagger }\left( x\right)
A^{\left( 1\right) }\hat{\Psi}\left( x\right) $ is now express in $\left( 
\mathfrak{p},\mathfrak{q}\right) $-space. Moreover, we express $\hat{\Psi}%
\left( x\right) $ in terms of the third quantization annihilation field
operator, namely, Eq. (\ref{q-decomp}). Now using the expression of $\hat{A}%
^{\left( 1\right) }$ in terms of the $\mathfrak{p}$ and $\mathfrak{q}$ field
variables of QFT, we write%
\[
\hat{A}=\sum\limits_{\mathfrak{p},\mathfrak{q}}A\left( \mathfrak{p},%
\mathfrak{q}\right) \ \hat{\Delta}\left( \mathfrak{p},\mathfrak{q}\right) 
\]%
where%
\begin{eqnarray}
A\left( \mathfrak{p},\mathfrak{q}\right) &=&\sum\limits_{\mathfrak{u}}e^{i2%
\mathfrak{u}\cdot \mathfrak{q}}\left\langle \mathfrak{p}+\mathfrak{u}%
\right\vert \hat{A}\left\vert \mathfrak{p}-\mathfrak{u}\right\rangle \\
\hat{\Delta}\left( \mathfrak{p},\mathfrak{q}\right) &=&\sum\limits_{%
\mathfrak{v}}e^{i2\mathfrak{p}\cdot \mathfrak{v}}\left\vert \mathfrak{q}+%
\mathfrak{v}\right\rangle \left\langle \mathfrak{q}-\mathfrak{v}\right\vert
\end{eqnarray}%
We have%
\begin{eqnarray*}
\dint \hat{\Psi}^{\dagger }\left( x\right) A^{\left( 1\right) }\hat{\Psi}%
\left( x\right) dx &=&\sum\limits_{\mathfrak{p},\mathfrak{q}}A\left( 
\mathfrak{p},\mathfrak{q}\right) \ \dint \hat{\Psi}^{\dagger }\left(
x\right) \hat{\Delta}\left( \mathfrak{p},\mathfrak{q}\right) \hat{\Psi}%
\left( x\right) dx \\
&=&=\sum\limits_{\mathfrak{p},\mathfrak{q}}A\left( \mathfrak{p},\mathfrak{q}%
\right) \ \sum\limits_{\mathfrak{v}}e^{i2\mathfrak{p}\cdot \mathfrak{v}%
}\left\langle \hat{\Psi}^{\dagger }\right\vert \left\vert \mathfrak{q}+%
\mathfrak{v}\right\rangle \left\langle \mathfrak{q}-\mathfrak{v}\right\vert
\left\vert \hat{\Psi}\left( x\right) \right\rangle \\
&=&\sum\limits_{\mathfrak{p},\mathfrak{q}}A\left( \mathfrak{p},\mathfrak{q}%
\right) \ \sum\limits_{\mathfrak{v}}e^{i2\mathfrak{p}\cdot \mathfrak{v}%
}\dsum\limits_{\mathfrak{q}^{\prime }}\check{\psi}^{\dagger }\left( 
\mathfrak{q}^{\prime },t\right) \left\langle \mathfrak{q}^{\prime
}\right\vert \left\vert \mathfrak{q}+\mathfrak{v}\right\rangle \left\langle 
\mathfrak{q}-\mathfrak{v}\right\vert \dsum\limits_{\mathfrak{q}^{\prime
\prime }}\check{\psi}\left( \mathfrak{q}^{\prime \prime },t\right)
\left\vert \mathfrak{q}^{\prime \prime }\right\rangle
\end{eqnarray*}%
and we end up with%
\[
A^{\left( 1\right) }=\left( N\right) ^{-1}\dsum\limits_{\mathfrak{p},%
\mathfrak{q,v}}A\left( \mathfrak{p},\mathfrak{q}\right) \left[ \ e^{2i%
\mathfrak{p},\mathfrak{v}}\ \check{\psi}^{\dagger }\left( \mathfrak{q}+%
\mathfrak{v}\right) \check{\psi}\left( \mathfrak{q}-\mathfrak{v}\right) %
\right] 
\]%
\qquad \qquad\ Then we can write $A^{\left( 1\right) }$ in terms of third
quantization annihilation and creation operators, $\check{\psi}\left( 
\mathfrak{q}\right) $ and $\check{\psi}^{\dagger }\left( \mathfrak{q}\right) 
$, as%
\[
A^{\left( 1\right) }=\left( N\right) ^{-1}\dsum\limits_{\mathfrak{p},%
\mathfrak{q,v}}A\left( \mathfrak{p},\mathfrak{q}\right) \left[ \ e^{2i%
\mathfrak{p},\mathfrak{v}}\ \check{\psi}^{\dagger }\left( \mathfrak{q}+%
\mathfrak{v}\right) \check{\psi}\left( \mathfrak{q}-\mathfrak{v}\right) %
\right] 
\]%
in complete analogy to the second quantize version \cite{bk:ref9}. Here
trace over discrete indices is implied. We can more meaningfully write,%
\[
A^{\left( 1\right) }=\left( N\right) ^{-1}\dsum\limits_{\mathfrak{p},%
\mathfrak{q}}A\left( \mathfrak{p},\mathfrak{q}\right) \ \check{f}\left( 
\mathfrak{p},\mathfrak{q}\right) 
\]%
with%
\[
\ \check{f}\left( \mathfrak{p},\mathfrak{q}\right) =\dsum\limits_{\mathfrak{v%
}}\ e^{2i\mathfrak{p},\mathfrak{v}}\ \check{\psi}^{\dagger }\left( \mathfrak{%
q}+\mathfrak{v}\right) \check{\psi}\left( \mathfrak{q}-\mathfrak{v}\right) 
\]%
where $\ \check{f}\left( \mathfrak{p},\mathfrak{q}\right) $ is analogous to
the Klimontovich second quantization operator \cite{bk:brittin} for
one-particle phase space distribution function. Here we are talking of QFT $%
\left( \mathfrak{p},\mathfrak{q}\right) $-phase space of course, not the
ordinary $\left( p.q\right) $-phase space. Thus, taking the average, we have%
\[
\left\langle \ \check{f}\left( \mathfrak{p},\mathfrak{q}\right)
\right\rangle =\dsum\limits_{\mathfrak{v}}\ e^{2i\mathfrak{p},\mathfrak{v}}\
\left\langle \check{\psi}^{\dagger }\left( \mathfrak{q}+\mathfrak{v}\right) 
\check{\psi}\left( \mathfrak{q}-\mathfrak{v}\right) \right\rangle 
\]%
We identify the correlation function, 
\[
\left\langle \check{\psi}^{\dagger }\left( \mathfrak{q}+\mathfrak{v}\right) 
\check{\psi}\left( \mathfrak{q}-\mathfrak{v}\right) \right\rangle
=G^{<}\left( \mathfrak{q}+\mathfrak{v,q}-\mathfrak{v}\right) 
\]%
where $G^{<}\left( \mathfrak{q}+\mathfrak{v,q}-\mathfrak{v}\right) $
corresponds to the Green function for particle distribution of second
quantize quantum transport theory. Since here it pertains to the third
quantization scheme, we will denote this by fractur style, $\mathfrak{G}%
^{<}\left( \mathfrak{q}+\mathfrak{v,q}-\mathfrak{v}\right) $, so that%
\begin{eqnarray*}
\left\langle \ \check{f}\left( \mathfrak{p},\mathfrak{q}\right)
\right\rangle &=&\dsum\limits_{\mathfrak{v}}\ e^{2i\mathfrak{p},\mathfrak{v}%
}\ \mathfrak{G}^{<}\left( \mathfrak{q}+\mathfrak{v,q}-\mathfrak{v}\right) \\
&=&f_{w}\left( \mathfrak{p},\mathfrak{q}\right)
\end{eqnarray*}%
Therefore, $f_{w}\left( \mathfrak{p},\mathfrak{q}\right) $ is the lattice
Weyl transform of $\mathfrak{G}^{<}$, where $f_{w}\left( \mathfrak{p},%
\mathfrak{q}\right) $ is the Wigner distribution function in the third
quantization, which is akin to the lattice Weyl transform of the Green
function, $G^{<}$, for particle distribution in \textit{second} quantized
quantum transport theory.

\section{Ballistic Transport Equation in the Third Quantization}

The ballistic transport equation for $f_{w}\left( \mathfrak{p},\mathfrak{q,t}%
\right) $ is given by the%
\begin{equation}
i\left( \frac{\partial }{\partial t}\right) f_{w}\left( \mathfrak{p},%
\mathfrak{q,t}\right) =LWT_{third}\left[ \hat{H},\hat{\rho}\right]
\label{dynamics}
\end{equation}%
which derives from the quantum Liouville equation,%
\[
i\left( \frac{\partial }{\partial t}\right) \hat{\rho}=\left[ \hat{H},\hat{%
\rho}\right] \equiv :\mathcal{L}\hat{\rho} 
\]%
where $\mathcal{L}$ is the Liouvillian operator and $LWT_{third}$ indicates
the lattice Weyl transform of the commutator of the Hamiltonian and density
matrix operator. Upon carrying out the $LWT_{third}$ operation, we obtain,%
\begin{eqnarray}
LWT_{third}\left[ \hat{H},\hat{\rho}\right] &=&\left[ \hat{H},\hat{\rho}%
\right] \left( \mathfrak{p},\mathfrak{q,t}\right)  \nonumber \\
&=&\cos \Lambda \left[ H\left( \mathfrak{p},\mathfrak{q,t}\right)
f_{w}\left( \mathfrak{p},\mathfrak{q,t}\right) -f_{w}\left( \mathfrak{p},%
\mathfrak{q,t}\right) H\left( \mathfrak{p},\mathfrak{q,t}\right) \right] 
\nonumber \\
&&-i\sin \Lambda \left\{ H\left( \mathfrak{p},\mathfrak{q,t}\right)
f_{w}\left( \mathfrak{p},\mathfrak{q,t}\right) +f_{w}\left( \mathfrak{p},%
\mathfrak{q,t}\right) H\left( \mathfrak{p},\mathfrak{q,t}\right) \right\}
\label{LWTw}
\end{eqnarray}%
where $\Lambda $ is the Poisson bracket operator [here we put $\hbar
\longrightarrow 1$],%
\[
\Lambda =\frac{1}{2}\left( \frac{\partial ^{H}}{\partial \mathfrak{p}}\cdot 
\frac{\partial ^{f_{w}}}{\partial \mathfrak{q}}-\frac{\partial ^{H}}{%
\partial \mathfrak{q}}\cdot \frac{\partial ^{f_{w}}}{\partial \mathfrak{p}}%
\right) 
\]%
The RHS of Eq. (\ref{LWTw}) can also be expressed as an integral which is
more suitable for numerical computation. These expressions simplify
considerably when the respective lattice Weyl transforms are scalar
functions. Then we have 
\begin{equation}
\left[ \hat{H},\hat{\rho}\right] \left( \mathfrak{p},\mathfrak{q,t}\right)
=-2i\sin \Lambda \left\{ H\left( \mathfrak{p},\mathfrak{q,t}\right) \
f_{w}\left( \mathfrak{p},\mathfrak{q,t}\right) \right\}  \label{diff}
\end{equation}%
and in integral form as%
\begin{equation}
\left[ \hat{H},\hat{\rho}\right] \left( \mathfrak{p},\mathfrak{q,t}\right) =%
\frac{1}{\left( 2\pi \right) ^{8}}\dint \mathcal{D}\mathfrak{p}^{\prime }%
\mathcal{D}\mathfrak{q}^{\prime }K_{H}^{s}\left( \mathfrak{p},\mathfrak{q};%
\mathfrak{p}^{\prime },\mathfrak{q}^{\prime }\right) \ f_{w}\left( \mathfrak{%
p}^{\prime },\mathfrak{q}^{\prime },t\right)  \label{int}
\end{equation}%
where 
\begin{eqnarray}
K_{H}^{s}\left( \mathfrak{p},\mathfrak{q};\mathfrak{p}^{\prime },\mathfrak{q}%
^{\prime }\right) &=&\dint \mathcal{D}\mathfrak{u}\mathcal{D}\mathfrak{v}%
\exp \left\{ \frac{i}{\hbar }\left[ \left( \mathfrak{p}-\mathfrak{p}^{\prime
}\right) \cdot \mathfrak{v}+\left( \mathfrak{q}-\mathfrak{q}^{\prime
}\right) \cdot \mathfrak{u}\right] \right\}  \nonumber \\
&&\times \left[ H\left( \mathfrak{p+}\frac{\mathfrak{u}}{2},\mathfrak{q-}%
\frac{\mathfrak{v}}{2},t\right) -H\left( \mathfrak{p-}\frac{\mathfrak{u}}{2},%
\mathfrak{q+}\frac{\mathfrak{v}}{2},t\right) \right]  \label{KsH}
\end{eqnarray}%
Thus, we have the ballistic transport equation in the third quantization
scheme given by%
\begin{equation}
i\left( \frac{\partial }{\partial t}\right) f_{w}\left( \mathfrak{p},%
\mathfrak{q,t}\right) =-2i\sin \Lambda \left\{ H\left( \mathfrak{p},%
\mathfrak{q,t}\right) \ f_{w}\left( \mathfrak{p},\mathfrak{q,t}\right)
\right\}  \label{diff_eq}
\end{equation}%
or%
\begin{equation}
i\left( \frac{\partial }{\partial t}\right) f_{w}\left( \mathfrak{p},%
\mathfrak{q,t}\right) =\frac{1}{\left( 2\pi \right) ^{8}}\dint \mathcal{D}%
\mathfrak{p}^{\prime }\mathcal{D}\mathfrak{q}^{\prime }K_{H}^{s}\left( 
\mathfrak{p},\mathfrak{q};\mathfrak{p}^{\prime },\mathfrak{q}^{\prime
}\right) \ f_{w}\left( \mathfrak{p}^{\prime },\mathfrak{q}^{\prime },t\right)
\label{int_eq}
\end{equation}%
where $K_{H}^{s}\left( \mathfrak{p},\mathfrak{q};\mathfrak{p}^{\prime },%
\mathfrak{q}^{\prime }\right) $ is given by Eq. (\ref{KsH}).

In the case of the numerical simulations of the \textit{second} quantization
scheme, the form of Eq. (\ref{int_eq}) has proved to be more fruitful for
simulating real nanodevices \cite{bk:ref11}. Equation (\ref{int_eq}) can be
generalized to matrix quantities, not just scalar, as well as to
energy-dependent quantities. In the above transport equations it is assumed
that the energy variable has been integrated out.

In what follows we give a general formulation of the dynamics of the
distribution functional in QFT in terms of the \textit{correlation functions}
of the third quantization annihilation, $\check{\psi}\left( \mathfrak{q}%
,t\right) $, and creation, $\check{\psi}^{\dagger }\left( \mathfrak{q}%
,t\right) $, operators. First, we have to cast all quantum operators,
especially one-body and two-body operators, in terms of $\check{\psi}%
^{\dagger }\left( \mathfrak{q},t\right) $ and $\check{\psi}\left( \mathfrak{q%
},t\right) $ and show that these have exactly the same form as that of the
second quantized many-body formulation \cite{bk:ref8} of CMP. This then
leads us to a parallel formulation of '\textit{third}' quantum superfield
theory of transport physics based on the quantum Liouville equation or von
Neumann equation.

\section{Hamiltonian Operators in the Third Quantization}

To construct the dynamics of the QFTdistribution in functional phase space,
we need to cast the Hamiltonian and all quantum operators in terms of the
third quantize annihilation, $\check{\psi}\left( \mathfrak{q}\right) $, and
creation operators, $\check{\psi}^{\dagger }\left( \mathfrak{q}\right) $,
where $\mathfrak{q}$ is the eigenvalue of the second quantize operator, $%
\hat{\psi}\left( x\right) $. Observe that we use the \textit{hat} for second
quantize operators, whereas we use \textit{inverted hat} for the third
quantize operators.

Let $\hat{A}_{1}$ and $\hat{A}_{2}$ denote one-body and two-body
quantum-mechanical operators, respectively. Following the usual prescription
of the many-body quantization procedure, we write $\hat{A}_{1}$ and $\hat{A}%
_{2}$ in terms of the field operators, $\Psi $ and its dual $\Psi ^{\dagger
} $ as%
\begin{eqnarray}
\hat{A}_{1} &=&\dint \tilde{\Psi}^{\dagger }\left( x\right) A_{1}\tilde{\Psi}%
\left( x\right)  \label{3-1} \\
\hat{A}_{2} &=&\diint \tilde{\Psi}^{\dagger }\left( x\right) \tilde{\Psi}%
^{\dagger }\left( x^{\prime }\right) A_{2}\tilde{\Psi}\left( x^{\prime
}\right) \tilde{\Psi}\left( x\right) \ dx\ dx^{\prime }  \label{3-2}
\end{eqnarray}%
As before, we decompose the field operators in terms eigenfunction $%
\left\vert \mathfrak{q}\right\rangle $ and its dual, $\left\langle \mathfrak{%
q}\right\vert $. We have, 
\begin{eqnarray}
\tilde{\Psi}\left( x\right) &=&\dsum\limits_{\mathfrak{q}}\check{\psi}\left( 
\mathfrak{q}\right) \left\vert \mathfrak{q}\right\rangle \\
\tilde{\Psi}^{\dagger }\left( x\right) &=&\dsum\limits_{\mathfrak{q}}\check{%
\psi}^{\dagger }\left( \mathfrak{q}\right) \left\langle \mathfrak{q}%
\right\vert  \label{pdecomp}
\end{eqnarray}%
where we have suppressed summation over other inherent discrete indices if
present in the theory. We have also not displayed the argument $x$ in the
eigenfunction basis.

Thus, $\check{\psi}\left( \mathfrak{q},t\right) $ and $\check{\psi}^{\dagger
}\left( \mathfrak{q},t\right) $, labeled by the second quantized $\mathfrak{q%
}$ field, annihilate and create (generate) second quantized fields,
indicated by $\left\vert \mathfrak{q}\right\rangle $ and $\left\langle 
\mathfrak{q}\right\vert $. These third quantized field operators, $\check{%
\psi}\left( \mathfrak{q},t\right) $ and $\check{\psi}^{\dagger }\left( 
\mathfrak{q},t\right) $, satisfy the equal-time commutation relation for
bosons and anticommutation for fermions, namely,%
\[
\left[ \check{\psi}\left( \mathfrak{q}\right) ,\check{\psi}^{\dagger }\left( 
\mathfrak{q}^{\prime }\right) \right] _{\pm }=\delta _{\mathfrak{q,q}%
^{\prime }} 
\]%
\[
\left[ \check{\psi}\left( \mathfrak{q}\right) ,\check{\psi}\left( \mathfrak{q%
}^{\prime }\right) \right] _{\pm }=0\text{,} 
\]%
where the $+$ subscript is for fermions and the $-$ is bosons. We desire to
express Eqs. (\ref{3-1}) and (\ref{3-2}) in terms of third quantized field
operators, $\check{\psi}\left( \mathfrak{q},t\right) $ and $\check{\psi}%
^{\dagger }\left( \mathfrak{q},t\right) $, labeled by $\mathfrak{q}$.

We also like to express $A_{1}$ and $A_{2}$ in terms of $\left( \mathfrak{p},%
\mathfrak{q}\right) $-functional phase space. Thus, we need to express $%
A_{1} $ and $A_{2}$ in terms of its lattice Weyl transforms and phase-space
point projectors. We have using the expression of $\hat{A}_{1}$ in terms of
the $\mathfrak{p}$ and $\mathfrak{q}$ field variables of QFT, we write%
\[
\hat{A}_{1}=\sum\limits_{\mathfrak{p},\mathfrak{q}}A\left( \mathfrak{p},%
\mathfrak{q}\right) \ \hat{\Delta}\left( \mathfrak{p},\mathfrak{q}\right) 
\]%
where%
\begin{eqnarray}
A\left( \mathfrak{p},\mathfrak{q}\right) &=&\sum\limits_{\mathfrak{u}}e^{i2%
\mathfrak{u}\cdot \mathfrak{q}}\left\langle \mathfrak{p}+\mathfrak{u}%
\right\vert \hat{A}\left\vert \mathfrak{p}-\mathfrak{u}\right\rangle \\
\hat{\Delta}\left( \mathfrak{p},\mathfrak{q}\right) &=&\sum\limits_{%
\mathfrak{v}}e^{i2\mathfrak{p}\cdot \mathfrak{v}}\left\vert \mathfrak{q}+%
\mathfrak{v}\right\rangle \left\langle \mathfrak{q}-\mathfrak{v}\right\vert
\end{eqnarray}%
We have%
\begin{eqnarray*}
\dint \hat{\Psi}^{\dagger }\left( x\right) A^{\left( 1\right) }\hat{\Psi}%
\left( x\right) dx &=&\sum\limits_{\mathfrak{p},\mathfrak{q}}A\left( 
\mathfrak{p},\mathfrak{q}\right) \ \dint \hat{\Psi}^{\dagger }\left(
x\right) \hat{\Delta}\left( \mathfrak{p},\mathfrak{q}\right) \hat{\Psi}%
\left( x\right) dx \\
&=&\sum\limits_{\mathfrak{p},\mathfrak{q}}A\left( \mathfrak{p},\mathfrak{q}%
\right) \ \sum\limits_{\mathfrak{v}}e^{i2\mathfrak{p}\cdot \mathfrak{v}%
}\left\langle \hat{\Psi}^{\dagger }\right\vert \left\vert \mathfrak{q}+%
\mathfrak{v}\right\rangle \left\langle \mathfrak{q}-\mathfrak{v}\right\vert
\left\vert \hat{\Psi}\left( x\right) \right\rangle
\end{eqnarray*}%
and substituting Eqs. (\ref{q-decomp}) and (\ref{p-decomp-2}), we end up with%
\[
A^{\left( 1\right) }=\left( N\right) ^{-1}\dsum\limits_{\mathfrak{p},%
\mathfrak{q,v}}A\left( \mathfrak{p},\mathfrak{q}\right) \left[ \ e^{2i%
\mathfrak{p},\mathfrak{v}}\ \check{\psi}^{\dagger }\left( \mathfrak{q}+%
\mathfrak{v}\right) \check{\psi}\left( \mathfrak{q}-\mathfrak{v}\right) %
\right] 
\]%
We can reduce to a more familiar form amenable to field theoretical
perturbation technique by integrating with respect to $\mathfrak{p}$ (here
summation means functional integration), and indicate the result by $%
W_{1}\left( 2\mathfrak{v},2\mathfrak{q}\right) $, i.e.,%
\begin{equation}
W_{1}\left( 2\mathfrak{v},2\mathfrak{q}\right) =\dsum\limits_{\mathfrak{p}%
}e^{2i\mathfrak{p},\mathfrak{v}}A\left( \mathfrak{p},\mathfrak{q}\right)
\label{only-p-sum}
\end{equation}%
which yields%
\[
A^{\left( 1\right) }=\dsum\limits_{\mathfrak{q,v}}W_{1}\left( 2\mathfrak{v},2%
\mathfrak{q}\right) \ \check{\psi}^{\dagger }\left( \mathfrak{q}+\mathfrak{v}%
\right) \check{\psi}\left( \mathfrak{q}-\mathfrak{v}\right) 
\]%
By transformation of variables, we end up with%
\[
A^{\left( 1\right) }=\dsum\limits_{\mathfrak{r,r}^{\prime }}W_{1}\left( 
\mathfrak{r}-\mathfrak{r}^{\prime },\mathfrak{r}+\mathfrak{r}^{\prime
}\right) \ \check{\psi}^{\dagger }\left( \mathfrak{r}\right) \check{\psi}%
\left( \mathfrak{r}^{\prime }\right) 
\]%
Using Eq. (\ref{lwt}) and (\ref{only-p-sum}), we have%
\[
W_{1}\left( 2\mathfrak{v},2\mathfrak{q}\right) =\dsum\limits_{\mathfrak{p}%
}e^{2i\mathfrak{p},\mathfrak{v}}A\left( \mathfrak{p},\mathfrak{q}\right) 
\]%
Substituting the expression for $A\left( \mathfrak{p},\mathfrak{q}\right) $
given by Eq. (\ref{lwt}), we have%
\begin{eqnarray*}
W_{1}\left( 2\mathfrak{v},2\mathfrak{q}\right) &=&\dsum\limits_{\mathfrak{p}%
}e^{2i\mathfrak{p},\mathfrak{v}}\sum\limits_{\mathfrak{v}^{\prime }}e^{i2%
\mathfrak{p}\cdot \mathfrak{v}^{\prime }}\left\langle \mathfrak{q}-\mathfrak{%
v}^{\prime }\right\vert A\left\vert \mathfrak{q}+\mathfrak{v}^{\prime
}\right\rangle \\
&=&\dsum\limits_{\mathfrak{p}}e^{2i\mathfrak{p},\mathfrak{v}}\sum\limits_{%
\mathfrak{v}^{\prime }}e^{-i2\mathfrak{p}\cdot \mathfrak{v}^{\prime
}}\left\langle \mathfrak{q}+\mathfrak{v}^{\prime }\right\vert A\left\vert 
\mathfrak{q}-\mathfrak{v}^{\prime }\right\rangle \\
&=&\dsum\limits_{\mathfrak{p,v}^{\prime }}e^{2i\mathfrak{p},\left( \mathfrak{%
v-v}^{\prime }\right) }\left\langle \mathfrak{q}+\mathfrak{v}^{\prime
}\right\vert A\left\vert \mathfrak{q}-\mathfrak{v}^{\prime }\right\rangle \\
&=&\dsum\limits_{\mathfrak{v}^{\prime }}\delta \left( \mathfrak{v-v}^{\prime
}\right) \left\langle \mathfrak{q}+\mathfrak{v}^{\prime }\right\vert
A\left\vert \mathfrak{q}-\mathfrak{v}^{\prime }\right\rangle \\
&=&\left\langle \mathfrak{q}+\mathfrak{v}\right\vert A\left\vert \mathfrak{q}%
-\mathfrak{v}\right\rangle =\left\langle \mathfrak{r}\right\vert A\left\vert 
\mathfrak{r}^{\prime }\right\rangle
\end{eqnarray*}%
Therefore we can write%
\begin{eqnarray}
A^{\left( 1\right) } &=&\dsum\limits_{\mathfrak{r,r}^{\prime }}W_{1}\left( 
\mathfrak{r},\mathfrak{r}^{\prime }\right) \ \check{\psi}^{\dagger }\left( 
\mathfrak{r}\right) \check{\psi}\left( \mathfrak{r}^{\prime }\right) 
\nonumber \\
&=&\dsum\limits_{1,2}V\left( 1,2\right) \ \check{\psi}^{\dagger }\left(
1\right) \check{\psi}\left( 2\right)  \label{2pointthird}
\end{eqnarray}%
Thus, we have succeeded in expressing the one-body Hamiltonian in terms of
the \textit{third} quantization creation, $\check{\psi}^{\dagger }$, and
annihilation, $\check{\psi}$, operators. This is formally the same
expression as in the conventional second quantization scheme.

Similarly, for the two-body operator we proceed as follows. We have%
\[
A^{\left( 2\right) }=\diint \Psi ^{\dagger }\left( x\right) \Psi ^{\dagger
}\left( x^{\prime }\right) A_{2}\Psi \left( x^{\prime }\right) \Psi \left(
x\right) dx^{\prime }dx 
\]%
which yields%
\begin{eqnarray*}
A^{\left( 2\right) } &=&\dsum\limits_{\mathfrak{p,p}^{\prime }\mathfrak{;q,q}%
^{\prime },\mathfrak{v}.\mathfrak{v}^{\prime }}A_{2}\left( \mathfrak{p,p}%
^{\prime }\mathfrak{;q,q}^{\prime }\right) \exp \left[ 2i\left( \mathfrak{%
p.v+p}^{\prime }\mathfrak{.v}^{\prime }\right) \right] \\
&&\times \check{\psi}^{\dagger }\left( \mathfrak{q}+\mathfrak{v}\right) 
\check{\psi}^{\dagger }\left( \mathfrak{q}^{\prime }+\mathfrak{v}^{\prime
}\right) \check{\psi}\left( \mathfrak{q}^{\prime }-\mathfrak{v}^{\prime
}\right) \check{\psi}\left( \mathfrak{q}-\mathfrak{v}\right)
\end{eqnarray*}%
where again summation over all other inherent discrete indices, like spins,
are suppressed to save space. Again, we can reduce to a more familiar form
amenable to field theoretical perturbation technique by integrating with
respect to $\mathfrak{p}$ and $\mathfrak{p}^{\prime }$ (here summation means
functional integration), and indicate the result by $W_{2}\left( 2\mathfrak{%
v,2v}^{\prime };\mathfrak{q,q}^{\prime }\right) $, i.e.,%
\[
W_{2}\left( 2\mathfrak{v,}2\mathfrak{v}^{\prime };\mathfrak{q,q}^{\prime
}\right) =\dsum\limits_{\mathfrak{p,p}^{\prime }}A_{2}\left( \mathfrak{p,p}%
^{\prime }\mathfrak{;q,q}^{\prime }\right) \exp \left[ 2i\left( \mathfrak{%
p.v+p}^{\prime }\mathfrak{.v}^{\prime }\right) \right] 
\]%
We have%
\begin{eqnarray*}
&&W_{2}\left( 2\mathfrak{v},2\mathfrak{q}\right) \\
&=&\dsum\limits_{\mathfrak{p}}e^{2i\mathfrak{p},\mathfrak{v}}\sum\limits_{%
\mathfrak{v}^{\prime \prime }}e^{i2\mathfrak{p}\cdot \mathfrak{v}^{\prime
\prime }}\left\langle \mathfrak{q}-\mathfrak{v}^{\prime \prime }\right\vert
A\left\vert \mathfrak{q}+\mathfrak{v}^{\prime \prime }\right\rangle
\dsum\limits_{\mathfrak{p}^{\prime }}e^{2i\mathfrak{p}^{\prime },\mathfrak{v}%
^{\prime }}\sum\limits_{\mathfrak{v}^{\prime \prime \prime }}e^{i2\mathfrak{p%
}^{\prime }\cdot \mathfrak{v}^{\prime \prime \prime }}\left\langle \mathfrak{%
q}^{\prime }-\mathfrak{v}^{\prime \prime \prime }\right\vert A\left\vert 
\mathfrak{q}^{\prime }+\mathfrak{v}^{\prime \prime \prime }\right\rangle \\
&=&\dsum\limits_{\mathfrak{p,v}^{\prime \prime }}e^{2i\mathfrak{p},\left( 
\mathfrak{v-v}^{\prime \prime }\right) }\dsum\limits_{\mathfrak{p}^{\prime },%
\mathfrak{v}^{\prime \prime \prime }}e^{2i\mathfrak{p}^{\prime },\left( 
\mathfrak{v}^{\prime }-\mathfrak{v}^{\prime \prime \prime }\right)
}\left\langle \mathfrak{q}+\mathfrak{v}^{\prime \prime }\right\vert
\left\langle \mathfrak{q}^{\prime }+\mathfrak{v}^{\prime \prime \prime
}\right\vert A\left\vert \mathfrak{q}^{\prime }-\mathfrak{v}^{\prime \prime
\prime }\right\rangle \left\vert \mathfrak{q}-\mathfrak{v}^{\prime \prime
}\right\rangle \\
&=&\left\langle \mathfrak{q}+\mathfrak{v}\right\vert \left\langle \mathfrak{q%
}^{\prime }+\mathfrak{v}^{\prime }\right\vert A\left\vert \mathfrak{q}%
^{\prime }-\mathfrak{v}^{\prime }\right\rangle \left\vert \mathfrak{q}-%
\mathfrak{v}\right\rangle =\left\langle \mathfrak{r}\right\vert \left\langle 
\mathfrak{r}^{\prime }\right\vert A\left\vert \mathfrak{r}^{\prime \prime
}\right\rangle \left\vert \mathfrak{r}^{\prime \prime \prime }\right\rangle
\end{eqnarray*}%
So we end up with the expression,%
\[
A^{\left( 2\right) }=\dsum\limits_{\mathfrak{r,r}^{\prime },r^{\prime \prime
},r^{\prime \prime \prime }}\left\langle \mathfrak{r}\right\vert
\left\langle \mathfrak{r}^{\prime }\right\vert A\left\vert \mathfrak{r}%
^{\prime \prime }\right\rangle \left\vert \mathfrak{r}^{\prime \prime \prime
}\right\rangle \check{\psi}^{\dagger }\left( \mathfrak{r}\right) \check{\psi}%
^{\dagger }\left( \mathfrak{r}^{\prime }\right) \check{\psi}\left( \mathfrak{%
r}^{\prime \prime }\right) \check{\psi}\left( \mathfrak{r}^{\prime \prime
\prime }\right) 
\]%
and can be written symbolically as,%
\begin{equation}
A^{\left( 2\right) }=\dsum\limits_{1,2,3,4}V\left( 1,2,3,4\right) \check{\psi%
}^{\dagger }\left( 1\right) \check{\psi}^{\dagger }\left( 2\right) \check{%
\psi}\left( 3\right) \check{\psi}\left( 4\right)  \label{4pointthird}
\end{equation}%
which is formally of the expression as those of second quantize operators
for four point potential \cite{bk:ref8}. Equations (\ref{2pointthird}) and (%
\ref{4pointthird}) allow us to write the Hamiltonian in terms of third
quantization, $\check{\psi}^{\dagger }$ and $\check{\psi}$, as,%
\[
\check{H}\mathfrak{=}\dsum\limits_{1,2}V\left( 1,2\right) \ \check{\psi}%
^{\dagger }\left( 1\right) \check{\psi}\left( 2\right)
+\dsum\limits_{1,2,3,4}V\left( 1,2,3,4\right) \check{\psi}^{\dagger }\left(
1\right) \check{\psi}^{\dagger }\left( 2\right) \check{\psi}\left( 3\right) 
\check{\psi}\left( 4\right) 
\]

The above developments allow us to map the quantum transport dynamics of CMP
to that of QFT, by essentially simple substitution of (a) second quantized
operators to third quantize field operators, (b) second quantize correlation
functions to third quantize correlation functions, and (c) crystal momentum, 
$p$ to $\mathfrak{p}$, as well as lattice position $q$ to $\mathfrak{q}$.
For example, we can still have a well-defined multi-component quantum
superfield operators in the third quantization, namely,%
\begin{equation}
\Psi \left( 1\right) =\left( 
\begin{array}{c}
\check{\psi}\left( 1\right) \\ 
\widetilde{\psi }^{\dagger }\left( 1\right) \\ 
\check{\psi}^{\dagger }\left( 1\right) \\ 
\widetilde{\psi }\left( 1\right)%
\end{array}%
\right) \equiv \left\{ \Psi _{\alpha }\left( 1\right) ,\text{ \ }\alpha
=1,2,3,4\right\} .  \label{rnc4.1}
\end{equation}%
Thus, we obtained a perfect match between the transport variables in CMP and
QFT. We summarize the identiications by the following Table,

\begin{center}
\begin{tabular}{|l|l|}
\hline
\multicolumn{2}{|l|}{\tiny \ \ \ \ \ \ \ \ \ \ \ \ \ \ \ \ \ \ \ \ \ \ \ \ \
\ \ \ \ \ \ \ \ \ \ \ \ \ \ \ \ \ \ \ \ \ \ \ \ \ \ \ \ \ \ \ \ Dynamical
Unification of CMP and QFT} \\ \hline
{\tiny \ \ \ \ \ \ \ \ \ \ \ \ \ \ \ \ \ \ \ \ \ \ CMP (in second
quantization)} & {\tiny \ \ \ \ \ \ \ \ \ \ \ \ \ \ \ \ \ \ \ \ \ \ \ \ \ \
\ \ QFT (in third quantization)} \\ \hline
{\tiny Bloch function, }$\left\vert p\right\rangle ${\tiny \ } & {\tiny %
Eigenfunction of \ 2nd quantization creation operator, }$\left\vert 
\mathfrak{p}\right\rangle $ \\ \hline
{\tiny Wannier function, }$\left\vert q\right\rangle $ & {\tiny %
Eigenfunction of \ 2nd quantization annihilation operator, }$\left\vert 
\mathfrak{q}\right\rangle $ \\ \hline
{\tiny 2nd quantization field operators, }$\hat{\psi}^{\dagger }\left(
q\right) ${\tiny \ and }$\hat{\psi}\left( q\right) $ & {\tiny 3rd
quantization quantum field operators, }$\check{\psi}^{\dagger }\left( 
\mathfrak{q}\right) ${\tiny \ and }$\check{\psi}\left( \mathfrak{q}\right) $
\\ \hline
{\tiny 2nd quantization field operators, }$\hat{\psi}^{\dagger }\left(
p\right) ${\tiny \ and }$\hat{\psi}\left( p\right) $ & {\tiny 3rd
quantization quantum field operators, }$\check{\psi}^{\dagger }\left( 
\mathfrak{p}\right) ${\tiny \ and }$\check{\psi}\left( \mathfrak{p}\right) $
\\ \hline
{\tiny Correlation functions, typified by }$G^{<}\left( q,q^{\prime
},t,t^{\prime }\right) $ & {\tiny Correlation functions, typified by }$%
\mathfrak{G}^{<}\left( \mathfrak{q},\mathfrak{q}^{\prime },t,t^{\prime
}\right) $ \\ \hline
{\tiny Momentum-position phase space, }$\left( p,q\right) $ & {\tiny %
"Momentum-position" functional phase space, }$\left( \mathfrak{p},\mathfrak{q%
}\right) $ \\ \hline
$\left\langle \left\langle 1\right\vert \right. \overline{S}\left( \infty
,-\infty \right) \left. \left\vert \rho _{eq}\right\rangle \right\rangle $ & 
$\left\langle \left\langle 1\right\vert \right. \mathcal{S}{\tiny \ }\left(
\infty ,-\infty \right) \left. \left\vert \rho _{eq}\right\rangle
\right\rangle $ \\ \hline
$=\exp \left[ \frac{i}{\hbar }W\right] ${\tiny generating function} & $=\exp %
\left[ \frac{i}{\hbar }\mathcal{W}\right] ${\tiny generating functional} \\ 
\hline
{\small Transport equations of correlation functions} & {\small Transport
equations of distribution functions} \\ \hline
\end{tabular}
\end{center}

\section{Quantum Superfield Transport Dynamics in Third Quantization}

We state here some results without going through similar formal procedure.
This similar procedure that can be followed is given in the authors book 
\cite{bk:ref8} and some references therein.

Here, we will simply complete the equation for $G^{<}$ of Eq. (\ref{diff_eq}%
) which is given only for ballistic transport in functional phase space $%
\left( \mathfrak{p},\mathfrak{q}\right) $. We have, upon following similar
procedure for second quantize superfield quantum transport formalism, the
full equation for the functional distribution correlation function, $%
\mathfrak{G}^{<}$, without Cooper pairings, as 
\begin{eqnarray}
i\left( \frac{\partial }{\partial t_{1}}+\frac{\partial }{\partial t_{2}}%
\right) \mathfrak{G}^{<}\left( 1,2\right) &=&\left[ H,\mathfrak{G}^{<}\right]
\left( 1,2\right)  \nonumber \\
&&+\left[ \Sigma ^{<},\func{Re}\mathfrak{G}^{r}\right] \left( 1,2\right) 
\nonumber \\
&&+\frac{i}{2}\left\{ \Sigma ^{<},A\right\} \left( 1,2\right)  \nonumber \\
&&-\frac{i}{2}\left[ \Gamma ,\mathfrak{G}^{<}\right] \left( 1,2\right)
\label{Gless-full-nopair}
\end{eqnarray}%
where$\left( 1,2\right) $ stands for $\left( \mathfrak{q}_{1},t_{1},%
\mathfrak{q}_{2},t_{2}\right) $. The quantities in Eq. (\ref%
{Gless-full-nopair}) are defined in the third quantization as follows 
\begin{eqnarray*}
H\left( 1,2\right) &=&\left[ H_{0}\left( 1,2\right) +V\left( 1,2\right) %
\right] \delta \left( 1-2\right) +\delta \left( t_{1}-t_{2}\right) \Sigma
^{HF}\left( \mathfrak{q}_{1},\mathfrak{q}_{2},t_{1}\right) +\func{Re}\Sigma
^{r}\left( 1,2\right) \\
\func{Re}\mathfrak{G}^{r}\left( 1,2\right) &=&\frac{-i}{2}\epsilon \left(
t_{1}-t_{2}\right) A\left( 1,2\right) \\
\func{Re}\Sigma ^{r}\left( 1,2\right) &=&\frac{-i}{2}\epsilon \left(
t_{1}-t_{2}\right) \Gamma \left( 1,2\right)
\end{eqnarray*}%
Upon taking the lattice Weyl transform, neglecting the Cooper pairing terms,
a simpler equation,%
\begin{eqnarray}
&&\frac{\partial }{\partial t}\mathfrak{G}^{<}\left( \mathfrak{p},E,%
\mathfrak{q},t\right)  \nonumber \\
&=&2\sin \left( \check{\Lambda}\right) \left[ H\left( \mathfrak{p},E,%
\mathfrak{q},t\right) \mathfrak{G}^{<}\left( \mathfrak{p},E,\mathfrak{q}%
,t\right) +\Sigma ^{<}\left( \mathfrak{p},E,\mathfrak{q},t\right) \func{Re}%
\mathfrak{G}^{r}\left( \mathfrak{p},E,\mathfrak{q},t\right) \right] 
\nonumber \\
&&+\cos \left( \check{\Lambda}\right) \left[ \Sigma ^{<}\left( \mathfrak{p}%
,E,\mathfrak{q},t\right) A\left( \mathfrak{p},E,\mathfrak{q},t\right)
-\Gamma \left( \mathfrak{p},E,\mathfrak{q},t\right) \mathfrak{G}^{<}\left( 
\mathfrak{p},E,\mathfrak{q},t\right) \right]  \label{LWTtransport}
\end{eqnarray}%
The interest in integrating Eq. (\ref{LWTtransport}) with respect to the
energy, $E$, is to obtain an equation for the distribution of quasiparticles,%
\[
f_{w}\left( \mathfrak{p},\mathfrak{q},t\right) =\dint dE\left( -i\right) 
\mathfrak{G}^{<}\left( \mathfrak{p},E,\mathfrak{q},t\right) 
\]%
Equation (\ref{LWTtransport}) can also be given in integral form \cite%
{bk:refA,bk:ref8}, this will not be given here, the readers are referred to
Ref. \cite{bk:refA,bk:ref8}.

\section{Concluding Remarks}

Using the phase-space formalism of quantum mechanics, there seems to be a
self-similarity of the mathematical structure, through an iterative
algorithm, in going from solid-state crystalline phase space $\left(
p,q\right) $, to second quantization scheme $\left( \mathfrak{p},\mathfrak{q}%
\right) $ phase-space, and further into the third quantization scheme,
brought by the 3rd creation, $\check{\psi}^{\dagger }\left( \mathfrak{q}%
,t\right) $, and 3rd annihilation, $\check{\psi}\left( \mathfrak{q},t\right) 
$, field operators whose eigenvalues will lead to higher quantization
schemes. This then will produce similar generating functional and
nonequilibrium superfield quantum transport equations. This suggests that
quantization appears to possess a hierarchical and 'self-similar'
mathematical framework.

This third quantization may have been interpreted as equivalent to
second-quantized coupling constants in Ref. \cite{goodstrom} in their
application of their own third quantization to cosmology. The advantages and
utility of the third quantization scheme in quantum field theory needs
further research and investigations, e.g. this may have applications in
quantum fluctuation theory of systems with many degrees of freedom (in the
successful use of coherent state representation in harmonic oscillators and
nonlinear optics) and in cosmology.

\begin{acknowledgement}
The author is thankful to Gibson Maglasang for help in providing relevant
references. The author is grateful for a PCIEERD-DOST 'Balik' Scientist
consulting grant at the Cebu Normal University, Department of Physics and
Mathematics. He is grateful to Allan Roy Elnar, to the CNU Dean of Arts and
Sciences, Dr. Milagros Greif, and to Gibson Maglasang for hosting his
'Balik' Scientist visit at CNU, which was a bit hindered by covid-19.
\end{acknowledgement}

\end{document}